\documentclass[twocolumn]{autart}    

\usepackage{graphicx}          
\usepackage[dvips]{epsfig}    

\usepackage{cite}

\usepackage{textcomp}
\usepackage{amsmath,amsfonts,amssymb,color}

\def\endpf{\hspace*{\fill}~$\square$\par\endtrivlist\unskip}

\usepackage{epsfig}
\usepackage{psfrag}
\usepackage{algorithm}
\usepackage{algpseudocode}
\usepackage{array}
\usepackage{epstopdf}
\usepackage{mathtools}
\usepackage{enumerate}
\usepackage{soul,color}
\usepackage{wrapfig}
\usepackage{nccmath}

\newtheorem{problem}{Problem}

\newtheorem{coro}{Corollary}
\newtheorem{lemma}{Lemma}
\newtheorem{remark}{Remark}
\newtheorem{definition}{Definition}
\newtheorem{proposition}{Proposition}
\newtheorem{assumption}{Assumption}

\usepackage[dvipsnames]{xcolor}
\usepackage{color}

\newcommand{\baike}[1]{\textcolor{black}{#1}}
\newcommand{\baikes}[1]{\textcolor{black}{#1}}

\newcommand{\bshee}[1]{\textcolor{black}{#1}}
\newcommand{\update}[1]{\textcolor{black}{#1}}
\newcommand{\bks}
[1]{\textcolor{black}{#1}}
\newcommand{\baishe}
[1]{\textcolor{black}{#1}}
\newcommand{\Journal}
[1]{\textcolor{black}{#1}}

\DeclareMathOperator{\fix}{fix}

\begin{document}

\begin{frontmatter}

\title{\baishe{Control} of  SIR Epidemics: Sacrificing
Optimality for Feasibility} 
\vspace{-5ex}                                                

\thanks[footnoteinfo]{Corresponding author: Lei Xin. This work is supported by the grant NSF-ECCS-\#2238388.}

\author[1]{Baike She}\ead{bshe6@gatech.edu},    
\author[2]{Lei Xin}\ead{\baishe{lxinshenqing@gmail.com}},               
\author[3]{Shreyas Sundaram}\ead{sundara2\@purdue.edu},  
\author[3]{Philip E. Pare\thanksref{footnoteinfo}}\ead{philpare@purdue.edu}

\address[1]{School of Electrical and Computer Engineering, Georgia Institute of Technology, Atlanta, GA, 30318, USA} 
\address[2]{\baishe{Department of Computer Science and Engineering, The Chinese University of Hong Kong, Hong Kong, China}} 
\address[3]{Elmore Family School of Electrical and Computer Engineering, Purdue University, West Lafayette, IN 47907, USA}  
\vspace{-2ex}    
          

\begin{abstract}                          
We study the impact of  parameter estimation and state measurement errors on a control framework 
 \baishe{for} optimally mitigating the spread of epidemics. We capture the epidemic spreading process using a susceptible-infected-removed (SIR) epidemic model and consider \update{isolation} as the control strategy. We use a \update{control} strategy to remove \update{(isolate) a portion of the infected population. Our goal is to maintain the daily infected population below a certain level, while minimizing the resource captured by the isolation rate.} Distinct from existing works on leveraging control strategies in epidemic spreading, we propose a parameter estimation strategy and further characterize the 
 parameter estimation error bound. In order to deal with uncertainties,
 we propose a robust \update{control} strategy by overestimating the seriousness of the epidemic and study the feasibility of the system. Compared to the optimal \update{control} strategy, we establish that the proposed strategy under parameter estimation and measurement errors will  sacrifice
 optimality  to 
 flatten the curve.
\end{abstract} 
\end{frontmatter}

\section{Introduction}
\vspace{-2ex}
Effective management of both resources and risks is crucial during an epidemic. In the context of the recent COVID-19 pandemic, researchers have focused on studying optimal control formulations to inform policy-making for pandemic mitigation~\cite{tsay2020modeling, perkins2020optimal, acemoglu2021optimal, morris2021optimal, martins2023n, van2023effective}. One study~\cite{tsay2020modeling} measured the social and economic costs based on the intensity of the social distancing intervention. Meanwhile, researchers in~\cite{perkins2020optimal} calibrated epidemic models using data from the USA to examine the impact of social distancing restrictions. Additionally,~\cite{acemoglu2021optimal} proposed a strategy that combines molecular and serology testing to aid in epidemic mitigation, while~\cite{morris2021optimal} discussed the vulnerability of optimal or near-optimal control strategies in the face of model uncertainties. Beyond optimal control approaches, researchers \baishe{have} explored model predictive control frameworks~\cite{kohler2020robust,carli2020model,zino2021analysis,she2022mpcepi} and other strategies~\cite{khadilkar2020optimising,scarabaggio2021nonpharmaceutical,mubarak2022individual} to generate optimal or sub-optimal policies for epidemic mitigation. For example,~\cite{scarabaggio2021nonpharmaceutical} utilized a transmission network to identify vaccination targets, while~\cite{bastani2021efficient} employed a reinforcement learning framework to aid in addressing the COVID-19 mitigation problem. Additionally, other studies have explored epidemic control and resource allocation in various  ways~\cite{bloem2009optimal, nowzari2016epidemics, di2017optimal,sharomi2017optimal, di2019optimal,liu2019bivirus,dangerfield2019resource,preciado2014epidemic_optimal,han2015data}.

\vspace{-2ex}
Most of the aforementioned research was built upon prior knowledge of epidemic model parameters. Nevertheless, works on epidemic modeling and prediction \cite{chowell2017fitting, baker2018mechanistic,wilke2020predicting} have demonstrated the difficulty in precisely modeling and predicting the behavior of epidemic spreading processes. Hence, it is challenging to obtain accurate model parameters when solving epidemic modeling and control problems. 
One common approach to tackle the problem is to leverage continuous-time models to capture a spreading process and use real-world spreading data to fine-tune the model parameters.
While leveraging continuous-time models to analyze and design mitigation strategies for an epidemic spreading process is widely accepted~\cite{nowzari2016epidemics, casella2020can,sontag2023explicit, olejarz2023optimal}, spreading data 
\baishe{is typically only available}
in a discrete format, such as hourly, daily, weekly, and/or monthly infected cases.
Thus, it is critical to investigate the error introduced by leveraging sampled data to estimate model parameters for continuous-time spreading dynamics.
The existing literature offers estimation algorithms to estimate parameters of continuous-time models from sampled data~\cite{kirshner2014unique,zhu2022performance}. However, these methods cannot be directly applied to our setting to obtain explicit error bounds of the estimated parameters due to the specialized structure of the model we \baishe{consider}, as well as \baishe{measurement error}.

\vspace{-1ex}
In this study, our focus lies in addressing optimal epidemic \update{control} problems by studying feasible optimal control frameworks under uncertainties.
To bridge the gap between using continuous-time models to study the spread and having discrete-format data, we consider leveraging sampled states from continuous-time models to estimate the model parameters. These \baishe{estimated} parameters will be used for control design. Specifically, we will study the impact of discretization on analyzing continuous-time epidemic spreading models.
Our primary objective is to propose an algorithm for estimating model parameters through sampling data, particularly by utilizing spreading data from the early stage\baishe{s} of a pandemic within the same population of interest. 
By incorporating the estimated model parameters and \baishe{noisy} measured states, we aim to enhance our proposed optimal control strategy developed in continuous-time. %
The ultimate goal is to generate a robust control strategy for epidemic mitigation by leveraging the measured states and estimated model parameters, instead of relying solely on accurate model parameters and states.

\vspace{-1ex}
\update{In particular, we consider an existing optimal control algorithm~\cite{acemoglu2021optimal}, which may fail under measurement and parameter estimation errors~\cite{morris2021optimal}. Building upon this optimal control strategy, we propose a novel robust control strategy that can handle these uncertainties, albeit at a higher control cost.}
Furthermore, we generate an optimality gap between the optimal control strategy with perfect information and our proposed robust strategy with inaccurate information. 
Specifically, we consider a \update{isolation} strategy~\cite{acemoglu2021optimal} as the control input variable, which involves the removal of the infected population from the infected group through uniform random sampling. This strategy resembles vaccination strategies that target the susceptible population within the mixed group~\cite{grundel2021coordinate}, making the isolation strategy another widely adopted approach for epidemic mitigation~\cite{casella2020can,acemoglu2021optimal,she2022mpcepi,olejarz2023optimal,she2023reverse}.

\vspace{-1ex}
In summary, we propose a model parameter estimation strategy and investigate the impact of sampling rate and state measurement error on the estimation error bound. Additionally, we define a robust control strategy for epidemic mitigation that takes into account the influence of uncertainties resulting from both parameter estimation and \baishe{noisy states}.
To summarize, our contributions are:
\vspace{-1ex}
\begin{itemize}
    \item We formulate the parameter estimation problem as a linear regression problem to estimate epidemic model parameters. Additionally, we establish an estimation error bound considering the sampling rate and measurement error.
    
    \item We introduce a robust \update{control} strategy by overestimating the seriousness of the spread. This approach enables us to adapt the optimal control strategy by leveraging the bounds of the estimated model parameters and measured states, ensuring the feasibility of the system.
\item Our work bridges the gap between parameter estimation for epidemic spreading and the theoretical analysis of the feasibility of optimal control strategies for epidemic mitigation~\cite{sontag2023explicit}.
\item 
We examine the impact of parameter and state uncertainties on the proposed optimal policy~\cite{acemoglu2021optimal} by studying the additional control cost required for feasibility.  
\end{itemize}

\vspace{-1ex}
\update{Note that this paper is an extension of our conference paper~\cite{she2022optimal}. In~\cite{she2022optimal}, 
we consider a non-zero lower bound on the control input, whereas in this work, the lower bound on the control input is zero, the same as that in~\cite{acemoglu2021optimal}. 
Furthermore, our robust control strategy will be exactly the same as the optimal control strategy, when we have accurate model and state estimation.
Additionally, building upon our conference paper~\cite{she2022optimal} where we mainly discuss the control problem,} in this work,
we introduce a parameter estimation method;
quantify the parameter estimation error bounds in Theorem~\ref{Error_bound} and Corollary~\ref{coro: range}; and study the impact of the parameter estimation error bounds and state measurement error bounds on the additional control cost in Theorem~\ref{thm: cost}.

\vspace{-1ex}
The paper is organized as follows. In Section~\ref{section2}, we introduce the parameter estimation problem, the optimal epidemic \update{control} problem, and the goal\baishe{s} of this work. Section~\ref{sec:2-5} presents a parameter estimation algorithm to estimate the epidemic model parameters. Additionally, we provide an upper bound for the error generated by the parameter estimation algorithm and further quantify the factors affecting the parameter estimation error bound.  
In Section~\ref{section3}, we introduce a robust control strategy to study the feasibility of the control problem under uncertainties. We characterize the additional control cost by comparing \baishe{the strategy} with the optimal \update{policy} generated \baishe{with} accurate models and \baishe{true} states.
Section~\ref{section4} illustrates the proposed model parameter estimation algorithm and control strategy through simulations. Finally, Section~\ref{section5} presents the conclusions and outlines potential future work.
\vspace{-2ex}
\subsection*{Notation}
\vspace{-2ex}
Vectors are considered as column vectors unless otherwise indicated. Let $\mathbb{R}$ and $\mathbb{R}_{>0}$ denote the sets of real numbers and positive real numbers, respectively. Let $\lambda_{\max}(\cdot)$ and $\lambda_{\min}(\cdot)$ represent the largest and smallest eigenvalues in magnitude of a given matrix, respectively. For a given matrix $A$, $A^\top$ denotes its conjugate transpose. The spectral norm and Frobenius norm of matrix  $A$ are denoted by $\|A\|$ and $\|A\|_{F}$, respectively. The closed $l_{2}$ ball in $d$-dimensional space with a center at $x_{0}$ and radius $r$ is
denoted by $\mathcal{B}_{d}(x_{0},r)\triangleq \{x\in \mathbb{R}^{d}:\|x-x_{0}\|\leq r\}$.
 \update{We use  $X^*(t^*)$ to represent the state generated under the optimal control strategy. We use $X(t)$ to represent the state generated by a control strategy, excluding the optimal control strategy. We use 
$\hat{X}(\hat{t})$ to represent the measured state with measurement noise.}

\vspace{-2.5ex}
\section{Problem Formulation}
\vspace{-2ex}
\label{section2}
In this section, we introduce the epidemic spreading model and formulate the optimal resource allocation problem for epidemic mitigation. We first present the continuous-time model for analysis and control design. Subsequently, we introduce its corresponding discrete-time model for parameter estimation and error analysis, acknowledging the reality of having access only to sampled data. Our objective is to propose a potential approach for policymakers to implement a control strategy by estimating a spreading model. This approach aims to mitigate an epidemic using estimated model parameters and measured states.


\vspace{-2ex}
\subsection{\update{Isolation as the Control Strategy}}
\vspace{-2ex}
In this section, we  present the model for the epidemic mitigation problem. We consider the  closed-loop \baishe{susceptible-infected-recovered/removed} ($SIR$) model
\begin{subequations}
\vspace{-2.1ex}
\label{Eq: Con_Dynamics}
\begin{alignat}{3}
 \frac{dS(t)}{dt} &= -\beta S(t)I(t), \label{eq:S_u}\\
   \frac{dI(t)}{dt} &= \beta S(t)I(t) -(\gamma+u(t)) I(t),\label{eq:I_u}\\
    \frac{dR(t)}{dt} &= (\gamma+u(t)) I(t).\label{eq:R_u}
\vspace{-1.5ex}
\end{alignat}
\end{subequations}
We use \baishe{$S(t)$, $I(t)$, and $R(t)$} to represent the susceptible, infected, and removed proportion\baishe{s} of the population of interest, respectively. Note that $S(t)$, $I(t)$, $R(t)\in[0,1]$ and $S(t)+I(t)+R(t)=1$,  for all~$t\geq0$. The parameters $\beta$ and $\gamma$ represent the time-invariant transmission rate and removal rate, respectively, and the control input $u(t)$ captures the
isolation rate from the \update{isolation strategy. 
The \update{isolation} strategy can be achieved by uniformly randomly sampling $\bar{u}\in(0,1]$ of the proportion of the entire population (assuming the population is sufficiently large). Then, we can utilize quarantine resources to isolate/remove $u(t)\times100\%$ of the infected population from the infected group, where $u(t)\in [0,\bar{u}]$. 
We assume there is no sampling bias or delay in isolation, and that the process is completely accurate. This setting assumes that we have sufficient resources to monitor the severity of the spread by catching a proportion of the infected population. However, in order to change the spread, it is necessary to prevent the detected infected individuals from spreading the virus, which requires an isolation process. The strength of the isolation process is captured by the isolation rate $u(t)$. Note that $u(t)\in [0,\bar{u}]$, ensuring that the isolation rate does not exceed $\bar{u}$.
}

\vspace{-1ex}
Note that when $u(t)=0$, the system in \eqref{Eq: Con_Dynamics} \baishe{reduces to} the classic $SIR$ model \cite{kermack1927_sir}. Additionally, we use \textit{infection level} to describe the number of the infected proportion $I(t)$, for all~$t\geq0$. 
In this work, we assume the removal rate $\gamma$ captures any processes that separate the detected infected group from the entire population.
These processes may include recovery, hospitalization, death, etc.
We define mitigation as the act of keeping the infection level below a certain threshold through the implementation of the isolation strategy.
Through denoting 
\vspace{-1.1ex}
\begin{align}
\label{Eq_SIR_States}
    x(t)=[S(t) \ \ I (t) \ \ R(t)]^{\top},
\end{align}we represent the system in~\eqref{Eq: Con_Dynamics} via the  compact form $\dot x(t)=f(x(t),u(t))$.
\vspace{-1.4ex}
\subsection{Parameter Estimation}
\vspace{-2ex}
For real-world disease spread, the parameters for the spreading model in~\eqref{Eq: Con_Dynamics} are usually unknown. In addition, we can only access measured states by sampling the dynamics.
Hence, we first use
Euler's method with a sample step size of $h\in \mathbb{R}_{>0}$, to rewrite the system model in~\eqref{Eq: Con_Dynamics} 
as the following set of difference equations: 
\begin{subequations}\label{Eq:dt system}
\vspace{-2ex}
\begin{alignat}{3}
    S(t+h) &= S(t) - h\beta  S(t)I(t)+e_{S}(t), \label{eq:estimateS}\\
    I(t+h) &= I(t)+ h\beta S(t)I(t)
    -h(\gamma+u(t))  I(t)+e_{I}(t),\label{eq:estimateI}\\
    R(t+h) &=R(t) + h(\gamma+u(t)) I(t)+e_{R}(t),
    \label{eq:estimateR}
\end{alignat}
\end{subequations}
where $e_{S}(t)$, $e_{I}(t)$, $e_{R}(t)$ capture the discretization error. We will estimate the model parameters $\beta$ and $\gamma$ by leveraging~\eqref{eq:estimateS}-\eqref{eq:estimateR}, based on linear regression~\cite{draper1998applied}.  

\vspace{-1ex}
\baishe{The} existing literature \baishe{on} control frameworks for epidemic mitigation \baishe{estimated} model parameters through numerical optimization methods~\cite{roosa2019assessing,zimmer2017likelihood,kohler2020robust}. However, 
few works proposed rigorous analyses for the control design (optimality gap and feasibility) under the impact of both parameter estimation error and measurement error~\cite{tsay2020modeling,kohler2020robust,morato2020optimal,peni2020nonlinear,carli2020model,grundel2020much,sereno2021model,sontag2023explicit}. 
In \baishe{this} work, we discuss 
the impact of parameter estimation for $\beta$ and $\gamma$, and \baishe{noisy state measurements,} on the optimal control design.
\vspace{-2ex}
\subsection{Optimal \update{Control} Problem}
\vspace{-2ex}
In this section, we introduce the  optimal control framework.
Consider the system formulated in \eqref{Eq: Con_Dynamics}.  The objective of the control problem is to optimally allocate \update{isolation} resources during the pandemic in order to maintain the daily infected population at or below a desired infection threshold. 
In this work, we consider controlling the epidemic by minimizing the total \update{isolation resources} during the epidemic through the cost function 
\vspace{-2ex}
\begin{equation} 
\vspace{-2ex}
\label{eq:cost_function}
J(u(t))=\int_{0}^{+\infty}u(t)dt,
\end{equation}
\update{where the total isolation resources, such as the capacity of the quarantine centers we use, are proportional to the cumulative isolation rate $J(u(t))$.}
Hence, in order to obtain the isolation strategy that minimizes \update{the cumulative isolation rate} during the epidemic spreading process while ensuring that the fraction of infected 
individuals  
remains below a desired threshold, we formulate the following optimization problem, 
\begin{subequations}
\vspace{-2ex}
\label{eq:prob}
\begin{align}
&\min_{u(t),\update{ t\geq 0}}  \, \, J(u(t)) \\
&\text{s.t.}  \, \, 
\dot{x}(t)=f(x(t),u(t)),  \\ 
\label{eq:constraints}
&0 \leq I(t) \leq \bar{I}, 
\update{0} \leq {u}(t) \leq \bar{u}, 
\text{for all}~t \in [0, +\infty).
\end{align}
\end{subequations}
The control input constraints \update{zero} and $\bar{u}$ define the lower and upper bounds on the \update{isolation} rates, respectively. We have that $\overline{u}\in \mathbb{R}_{\geq0}$. With accurate model parameters \baikes{and states}, we can compute the closed-form solution to the optimal control problem in~\eqref{eq:prob} \cite{acemoglu2021optimal}. However, in reality, optimal control of real-time epidemic mitigation is challenging, due to uncertainties from model parameters and states~\cite{morris2021optimal}. \baishe{According to the analysis in~\cite{morris2021optimal}}, \baishe{following the optimal control strategy} with these uncertainties may cause the infected population to exceed the infection threshold significantly, i.e., to cause huge outbreaks. 
Hence, 
our robust strategy aims to prioritize feasibility of the control strategy over \baishe{the} optimality of the solution under uncertainties~\cite{van2023effective}.
\vspace{-1.5ex}
\subsection{Problem Statement} 
\vspace{-2ex}
We focus on the theoretical analysis of the optimal control for the epidemic mitigation problem in~\eqref{eq:prob}, considering the impact of parameter and state uncertainties. We first propose a new algorithm to estimate the epidemic spreading model parameters through sampled measured states. 
Next, we study optimal control strategies for the problem defined in~\eqref{eq:prob} to derive a robust \update{isolation} strategy based on the estimated model parameters and measured states.
We explore the additional control cost by comparing the \update{cumulative isolation rate} generated from the proposed robust control strategy with the \update{cumulative isolation rate} conducted under the optimal \update{control} strategy. Our objective is to demonstrate the effectiveness of the robust \update{control} strategy, which involves overestimating the seriousness of the epidemic. 

\vspace{-1ex}
To elaborate on these goals, we will address the following problems in this work.
\vspace{-1ex}
\begin{problem} [Parameter Estimation]
\label{Prob-1}
    How can we develop an algorithm to estimate the model parameters $\beta$ and $\gamma$ in~\eqref{Eq:dt system} through measured states. Specifically, can we describe the estimation error bound in terms of the sample step size $h$ and the measurement error? 
\end{problem} 
\vspace{-1ex}
\begin{problem} [Feasibility Vs. Optimality]
    \label{Prob-2}
    In epidemic \update{control} problems, what strategies can we employ to ensure the feasibility of the problem in ~\eqref{eq:prob}? 
\end{problem}
\vspace{-1ex}
\begin{problem} [Additional Cost]
     \label{Prob-3}
    What is the additional control cost associated with leveraging the estimated parameters and measured states? How does the error bound of the estimated parameters affect the additional control cost?
\end{problem}
\vspace{-1ex}
\begin{problem} [Benefits and Limitations]
     \label{Prob-4}
    Compared to the optimal control strategy developed from accurate spreading models and states, what are the benefits and potential limitations of our proposed robust control strategy?
\end{problem}

\vspace{-1ex}
We will answer Problem~1 in Section~\ref{sec:2-5}, and Problems~2-4 in Section~\ref{section3}. 
We will also illustrate these results through simulations in \baishe{Section~\ref{section4}}.

\vspace{-2.5ex}
\section{Parameter Estimation}
\vspace{-2.5ex}
\label{sec:2-5}
In this section, we study the parameter estimation problem via the 
discrete-time spreading model in~\eqref{Eq:dt system}.
Recall that in real-world epidemic spreading processes, we can only access sampled data. 
Hence,
we propose an estimation strategy using linear regression to 
estimate the epidemic parameters in~\eqref{Eq:dt system}. Furthermore, we investigate the error bound of the estimated parameters through the sampling step size, which corresponds to the way of recording and reporting spreading data, such as hourly, weekly, or even monthly.

\vspace{-1ex}
It can be observed that the 
closed-loop epidemic spreading 
dynamics in~\eqref{Eq:dt system} are nonlinear with respect to the dynamic states. 
However, the equations are linear with respect to the spreading parameters $\beta$ and $\gamma$~\cite{mei2017epidemics_review}. 
Hence, it is reasonable that we propose a linear regression method to estimate the model parameters.
\baishe{We estimate the parameters $\beta$ and $\gamma$ by leveraging~\eqref{eq:estimateI}, which includes both parameters.}
Thus, for the parameter estimation process, 
we rearrange \eqref{eq:estimateI} in the following form:
\vspace{-3ex}
\begin{equation} 
\begin{aligned} 
I(t+h)-I(t)+hu(t)I(t)=\begin{bmatrix}\beta&\gamma\end{bmatrix}\begin{bmatrix}hS(t)I(t)\\
-hI(t)\end{bmatrix}+e_{I}(t).
\end{aligned}
\end{equation}
For $t\geq 0$, let the measured susceptible and infected states be $\hat{S}(t)=S(t)+v_{S}(t)$ and $\hat{I}(t)=I(t)+v_{I}(t)$, respectively,
 where $v_S(t)$ and $v_I(t)$, $t\geq0$, are real-valued measurement error\baishe{s}.
We further define $\Theta \triangleq \begin{bmatrix}\beta&\gamma\end{bmatrix}\in \mathbb{R}_{>0}^{1\times2}$, $l(t)\triangleq \hat{I}(t+h)-\hat{I}(t)+hu(t)\hat{I}(t)\in \mathbb{R}$, and $z(t)\triangleq\begin{bmatrix}\hat{S}(t)\hat{I}(t)\\
-\hat{I}(t)\end{bmatrix}\in \mathbb{R}^{2\times1}$.  Then, we have
\vspace{-4ex}
\begin{equation} 
\label{touse1}
\begin{aligned} 
\underbrace{
\hat{I}(t+h)-\hat{I}(t)+hu(t)\hat{I}(t)}_{l(t)}&=\underbrace{\begin{bmatrix}\beta&\gamma\end{bmatrix}}_{\Theta} \underbrace{\begin{bmatrix}\hat{S}(t)\hat{I}(t)\\
-\hat{I}(t)\end{bmatrix}}_{z(t)}h\\
&+e_{I}(t)+w(t),\\
\end{aligned}
\vspace{-2ex}
\end{equation}
where the aggregated measurement error is 
\vspace{-2ex}
\begin{equation} 
\begin{aligned} 
w(t)&\triangleq v_{I}(t+h)-v_{I}(t)+hu(t)v_{I}(t)\\
&+\begin{bmatrix}\beta&\gamma\end{bmatrix}\begin{bmatrix}-\hat{S}(t)v_{I}(t)-v_{S}(t)v_{I}(t)-v_{S}(t)\hat{I}(t)\\
v_{I}(t)\end{bmatrix}h.
\end{aligned}
\end{equation}
For all $t\geq0$, we can write \eqref{touse1} more compactly as
\vspace{-1ex}
\begin{equation} 
\begin{aligned} 
\label{eq:touse1_cp}
l(t)=\Theta z(t)h+e_{I}(t)+w(t).
\end{aligned}
\vspace{-1ex}
\end{equation}
For any fixed $i,j\geq 0$\baishe{,} where $i \neq j$, we further
define the batch matrices 
\vspace{-1.5ex}
\begin{equation} 
\label{eq:batch}
\begin{aligned} 
&L=\begin{bmatrix} l(i)&l(j)\end{bmatrix}\in \mathbb{R}^{1\times 2}, 
&Z=\begin{bmatrix} z(i)&z(j)\end{bmatrix}\in \mathbb{R}^{2 \times 2},\\
&E=\begin{bmatrix} e_{I}(i)&e_{I}(j)\end{bmatrix}\in \mathbb{R}^{1 \times 2},
&W=\begin{bmatrix} w(i)&w(j)\end{bmatrix}\in \mathbb{R}^{1 \times 2}.\\
\end{aligned}
\vspace{-1.5ex}
\end{equation}
Following \eqref{eq:touse1_cp}, we have 
\begin{equation} 
\vspace{-1ex}
\label{Eq: Compact}
\begin{aligned} 
L=\Theta Zh+E+W.
\end{aligned}
\vspace{-1ex}
\end{equation}
Based on~\eqref{Eq: Compact}, in order to estimate the transmission rate $\beta$ and removal rate $\gamma$ represented by $\Theta$, we can solve the following least squares estimation problem
\vspace{-1ex}
\begin{equation}
\begin{aligned}
\label{eq: optimization}
\mathop{\min}_{\tilde{\Theta}\in \mathbb{R}^{1\times 2}} \{\|L-\tilde{\Theta}Zh\|^{2}_{F}\},
\end{aligned}
\vspace{-2.5ex}
\end{equation}
where $\tilde{\Theta}$ represents the possible solutions (the transmission rate and removal rate) from the parameter space.
We denote the solution of the optimization problem in~\eqref{eq: optimization} as $\hat{\Theta}\triangleq \begin{bmatrix}\hat{\beta}&\hat{\gamma}\end{bmatrix}$. \baishe{Assuming that~$ZZ^\top$ is invertible, the} closed-form of~$\hat{\Theta}$ is given by
\vspace{-1.5ex}
\begin{equation} 
\label{eq:para_est}
\begin{aligned}
\hat{\Theta}=\frac{LZ^{^\top}(ZZ^\top)^{-1}}{h}.
\end{aligned}
\end{equation}
Consequently, leveraging \eqref{Eq: Compact}, the estimation error is then given by
\vspace{-2ex}
\begin{equation} 
\label{error}
\begin{aligned}
\|\hat{\Theta}-\Theta\|&=\left\|\frac{EZ^\top(ZZ^\top)^{-1}}{h}+\frac{WZ^\top(ZZ^\top)^{-1}}{h}\right \|.\\
\end{aligned}
\end{equation}
Equation~\eqref{error} indicates that the estimation error \baishe{depends on the sample step size $h$. Thus,} 
in order to design a robust \update{control} strategy,
we further capture the impact of the sample step size $h$ and the measurement errors $v_I$ and $v_S$ on the estimation error bound  $\|\hat{\Theta}-\Theta\|$.
We introduce Lemma~\ref{dt_error}~\cite[Thm. 3.1]{malik2013discretization}, 
which helps bound the discretization error using Euler's method.
\begin{lemma}
\vspace{-2ex}
\label{dt_error}
For any $t\geq 0$, suppose that the system in~\eqref{Eq:dt system} is supplied with input using zero order hold with a sample step size $h>0$, i.e., $u(k)=u(t)$ for $k\in [t,t+h)$.  If the function $f$ is locally Lipschitz over a closed ball of radius $r$ around $x(t)$ with Lipschitz constant $\zeta$, i.e.,
\vspace{-1ex}
\begin{equation} 
\begin{aligned}
\|f(x,u(t))-f(y,u(t))\|\leq \zeta\|x-y\|, \forall x,y\in \mathcal{B}_{3}(x(t),r),
\end{aligned}
\vspace{-1ex}
\end{equation}
then the discretization error in \eqref{Eq:dt system} satisfies
\vspace{-2ex}
\begin{equation} 
\begin{aligned}
\left\|\begin{bmatrix}e_{S}(t) & e_{I}(t)& e_{R}(t)\end{bmatrix}\right\| \leq \frac{h^2\zeta \|f(x(t),u(t))\|}{1-\zeta h},
\end{aligned}
\vspace{-2ex}
\end{equation}
supposing that $h\leq h^*$, where $h^*$ is a threshold that depends on $\zeta,r, \|f(x(t),u(t))\| $.
\end{lemma}
\vspace{-1.5ex}
\begin{remark}
     Lemma~\ref{dt_error} bounds the discretization error through the Lipschitz constant $\zeta$, the change of the epidemic spreading $\|f(x(t),u(t))\|$, \baishe{and,} most importantly, the sample step size $h$. 
 The Lipschitz constant $\zeta$ along with $\|f(x(t),u(t))\|$ 
 capture the speed of the change of the spread, e.g., slowly or rapidly. In addition, the sampling step size $h$ is determined by data reporting and the reporting interval we choose to estimate the parameters.
\end{remark}
\vspace{-1.5ex}
Based on Lemma~\ref{dt_error}, 
We derive the following theorem to further characterize the parameter estimation error based on the spreading behavior \baishe{and the length of the sampling interval}. 
Notice that in the following theorem, we let $u_{\max}=\max\{|u_{i}|,|u_{j}|\}$, $x_{\max}=\max\{\|x(i)\|, \|x(j)\|\}$, $v_{\max}=\max\{|v_{I}(i)|,|v_{I}(i+h)|,|v_{I}(j)|,|v_{I}(j+h)|,|v_{S}(i)|,|v_{S}(i+h)|,|v_{S}(j)|,|v_{S}(j+h)|\}$, and $f_{\max}=\max\{\|f(x(i),u(i))\|, \|f(x(j),u(j))\|\}$.
\vspace{-1.5ex}
\begin{thm}
\label{Error_bound}
Suppose that $u(t)=u(i)$ for $t\in [i,i+h)$ and $u(t)=u(j)$ for $t\in [j,j+h)$, i.e., the input is applied via \baishe{a} zero order hold.  
Fix any $r>0$. Supposing that the sampling step size $h<h^*$, where $h^*$ depends on $\zeta,r,\|f(x(i),u(i))\|, \|f(x(j),u(j))\|$,  then the error in \eqref{error} satisfies
\begin{equation*} 
\begin{aligned}
\|\hat{\Theta}-\Theta\|&\leq b, \label{bound}
\end{aligned}
\vspace{-1ex}
\end{equation*}
where 
\vspace{-1ex}
\begin{align}
\vspace{-2ex}
\label{eq:e_bound}
\nonumber
b&=\frac{2h\zeta f_{\max}}{\sqrt{\lambda_{\min}(ZZ^\top)}(1-\zeta h)}+\frac{4v_{\max}}{h\sqrt{\lambda_{\min}(ZZ^\top)}}\\
&+\frac{v_{\max} c}{\sqrt{\lambda_{\min}(ZZ^\top)}},
\\
\zeta&=4\beta(x_{\max}+r)+2u_{\max}+2\gamma, \nonumber
\vspace{-2ex}
\end{align}
\vspace{-3ex}
\begin{equation} 
\begin{aligned}
c&=2u_{\max}+2\gamma+\beta(\hat{S}(i)+\hat{S}(j)+2v_{\max}+\hat{I}(i)+\hat{I}(j)).
\end{aligned}
\end{equation}
\end{thm}
\vspace{-3ex}
\begin{pf}
Consider the error in \eqref{error}, based on the triangle inequality and Cauchy–Schwarz inequality, we have
\begin{equation} \label{touse0}
\begin{aligned}
\|\hat{\Theta}-\Theta\|\leq \frac{\|E\|\|Z^\top(ZZ^\top)^{-1}\|}{h}+\frac{\|W\|\|Z^\top(ZZ^\top)^{-1}\|}{h}.
\end{aligned}
\end{equation}
We consider the term $\frac{\|E\|}{h}$ first. Recall from~\eqref{Eq_SIR_States} that $x(t)=[S(t) \ \ I (t) \ \ R(t)]^{\top}$. The Jacobian of $f(\cdot,u(i))$ is given by
\begin{equation*} 
\begin{aligned}
\mathbf{J}(x(t))=\begin{bmatrix}-\beta I(t) & -\beta S(t) & 0\\
\beta I(t) & \beta S(t)-\gamma-u(i)&0\\
0 & \gamma+u(i)&0\\
\end{bmatrix},
\end{aligned}
\end{equation*}
from which we have
\vspace{-1.5ex}
\begin{equation} 
\label{Jacobian}
\begin{aligned}
\|\mathbf{J}(x(t))\|\leq \|\mathbf{J}(x(t))\|_{F}\leq 2\beta I(t)+2\beta S(t)+2u(i)+2\gamma. 
\end{aligned}
\end{equation}
Fix any $r>0$, for all $x(t) \in \mathcal{B}_{3}(x(i),r)$, we have
\vspace{-1.5ex}
\begin{equation*} 
\begin{aligned}
\|x(t)-x(i)\|\leq r \Rightarrow \|x(t)\|\leq \|x(i)\|+r,
\end{aligned}
\vspace{-1.5ex}
\end{equation*}
which implies
\vspace{-1.5ex}
\begin{align*}
|S(t)|\leq \|x(i)\|+r, \textnormal{ and } |I(t)|\leq \|x(i)\|+r,
\vspace{-1.5ex}
\end{align*}
since both $S(t)$ and $I(t)$ are components of $x(t)$.
Thus, leveraging \eqref{Jacobian}, for all $x(t)\in \mathcal{B}_{3}(x(i),r)$, we have
\begin{equation*} 
\begin{aligned}
\|\mathbf{J}(x(t))\|&\leq 2\beta(\|x(i)\|+r)\\
&+2\beta (\|x(i)\|+r)+2u(i)+2\gamma\\
&\leq 4\beta(x_{\max}+r)+2u_{\max}+2\gamma\triangleq\zeta, 
\end{aligned}
\vspace{-1.5ex}
\end{equation*}
where $\zeta$ is a Lipschitz constant when restricting the domain of $f(\cdot, u(i))$ to $B(x(i),r)$. Following a similar procedure, we have $\|\mathbf{J}(x(t))\|\leq \zeta$ for all $x(t)\in \mathcal{B}_{3}(x(j),r)$. Consequently, for sufficiently small $h$, we can apply Lemma \ref{dt_error} to obtain
\vspace{-1.5ex}
\begin{equation} \label{touse2}
\begin{aligned}
\frac{\|E\|}{h} \leq  \frac{|e_{I}(i)|+|e_{I}(j)|}{h}\leq \frac{2h\zeta f_{\max}}{1-\zeta h}.
\end{aligned}
\vspace{-3ex}
\end{equation}
Next, for the term $\frac{\|W\|}{h}$, we have
\vspace{-1.5ex}
\begin{equation}  \label{touse3}
\begin{aligned}
&\frac{\|W\|}{h}\leq \frac{|w(i)|+|w(j)|}{h}\leq \frac{4v_{\max}}{h}+\\
&v_{\max}(2u_{\max}+\beta(\hat{S}(i)+\hat{S}(j)\\
&\quad \quad\quad +2v_{\max}+\hat{I}(i)+\hat{I}(j))+2\gamma).
\end{aligned}
\vspace{-1.5ex}
\end{equation}
Finally, for the term $\|Z^\top(ZZ^\top)^{-1}\|$, we have
\vspace{-1.5ex}
\begin{equation}  \label{touse4}
\begin{aligned}
\|Z^\top(ZZ^\top)^{-1}\|&=\sqrt{\lambda_{\max}((ZZ^\top)^{-1}ZZ^\top(ZZ^\top)^{-1})}\\
&=\frac{1}{\sqrt{\lambda_{\min}(ZZ^\top)}}.
\end{aligned}
\end{equation}
The result follows by combining~\eqref{touse0} and \eqref{touse2}-\eqref{touse4}.
\end{pf}
\vspace{-1ex}
Theorem \ref{Error_bound} provides an error bound for estimating $\Theta$. The first term in~\eqref{eq:e_bound} is a quantity that captures the error due to the sampling rate $h$, and the second term and third term capture the error due to imperfect measurement $v_{\max}$. Note that the data-dependent term $\lambda_{\min}(ZZ^\top)$ does not depend on $h$. To gain more insights, suppose that for all $v_{\max}$, $h$, we have $\lambda_{\min}(ZZ^\top)>\lambda$ for some $\lambda>0$ for now (such that one can replace $\lambda_{\min}(ZZ^\top)$ by $\lambda$ in the error bound). We can see that a small sampling step size $h$ can drive the first term of the error bound arbitrarily small. However, a smaller $h$ will also cause the second term larger if we have a smaller signal to noise ratio on the measured states. To further reduce the second and third terms, one needs to reduce the measurement error $v_{\max}$ as well, i.e., small measurement error is critical \baishe{for} generating small parameter estimation error. 
\vspace{-1.5ex}
\begin{remark}
\label{Remark_Summary}
The mathematical analyses indicate that in order to achieve smaller errors in estimating the transmission rate $\beta$ and removal rate $\gamma$, it is crucial to use an appropriate data reporting interval depending on the measurement error and the rate of spread change. When dealing with a slowly changing spread and a high degree of measurement error, it is reasonable to estimate the model parameters using sample points with slower sampling rate (larger $h$), such as weekly or monthly data. Conversely, for a rapidly changing spread with minimal measurement error, faster sampling speed (smaller $h$), e.g., daily infection data, can yield relatively accurate parameter estimates. Therefore, 
Theorem~\ref{Error_bound} indicates that it is essential to consider multiple factors to obtain more accurate estimation results in practical applications. In the sequel (Remark \ref{remark_insight}), we will see that better estimates of the parameters lead to better performance of the control law (i.e., smaller optimality gap).
\end{remark}
\vspace{-2ex}
\begin{coro}
\label{coro: range}
Under the same conditions in Theorem \ref{Error_bound},  we have that
$
\hat{\beta}-b\leq \beta\leq \hat{\beta}+b \textnormal{ and, }
\hat{\gamma}-b\leq \gamma\leq \hat{\gamma}+b$.
\end{coro}
\vspace{-4ex}
\begin{pf}
We will only show the first inequality, as the proof for the second one is almost \baishe{identical}. When the bound in \baishe{Theorem~\ref{Error_bound}} holds, note that we have $|\beta-\hat{\beta}|=|\hat{\beta}-\beta|\leq \|\hat{\Theta}-\Theta\| \leq b$,
since $\hat{\beta}-\beta$ is a \baishe{scalar subvector of the vector} $\hat{\Theta}-\Theta$. Consequently, we have $-b\leq \beta-\hat{{\beta}} \leq b \Rightarrow \hat{\beta}-b\leq \beta\leq \hat{\beta}+b$. 
\end{pf}
\vspace{-1.5ex}
Corollary~\ref{coro: range}  provides bounds on the spreading parameters $\beta$ and $\gamma$. \baishe{Note that Theorem~\ref{Error_bound} considers estimating the model parameters through sampling data at two different time steps, which is common in estimating parameters for epidemic compartmental models~\cite{pare2018analysis}. Future work will also consider the connection between the estimation error bound and the increment of the sampling points.}
Next, we will present our robust control strategy for the epidemic mitigation problem, taking into account the estimation error bounds developed in this section. In summary, \baishe{Corollary~\ref{coro: range}, along with Theorem~\ref{Error_bound}, answers the parameter estimation problem given by Problem~1.}
\vspace{-3ex}
\section{\update{Control of Epidemic Spread}}
\vspace{-2.5ex}
\label{section3}
In this section, 
we first investigate a feasible control strategy for the problem defined in~\eqref{eq:prob} under the condition that the bounds of the 
model parameters $\beta$ and $\gamma$ are known. 
\vspace{-2ex}
Based on Corollary~\ref{coro: range}, we define that
\begin{align}
\label{eq:error_bounds}
\beta\in [\underbrace{\hat{\beta}-b}_{\hat{\beta}_{\min}},\underbrace{\hat{\beta}+b}_{\hat{\beta}_{\max}}],
\gamma \in [\underbrace{\hat{\gamma}-b}_{\hat{\gamma}_{\min}},\underbrace{\hat{\gamma}+b}_{\hat{\gamma}_{\max}}].
\vspace{-3ex}
\end{align}
Hence, we leverage the error bound generated through Corollary~\ref{coro: range} to design our robust \update{control} framework.
\baishe{We further assume that we know the bounds for the measured states ($\hat{S}(t)$ and $\hat{I}(t)$) and the true states ($S(t)$ and $I(t)$), \baishe{as follows}.}
\vspace{-2ex}
\begin{assumption}
\label{Assum_bounds}
The measured states $\hat{S}(t)$, $\hat{I}(t)$, and the true states $S(t)$, $I(t)$ follow
\vspace{-4ex}
\small
\begin{equation*} 
\begin{aligned}
\hat S(t),
S(t) \in[\hat{S}_{\min}(t),\hat{S}_{\max}(t)], \hat I(t),
I(t) \in[\hat{I}_{\min}(t),\hat{I}_{\max}(t)],
\end{aligned}
\vspace{-2ex}
\end{equation*}
\normalsize
\baishe{where $\hat{S}_{\min}(t)$ and $\hat{S}_{\max}(t)$ are the lower and upper bounds for the true susceptible state $S(t)$, respectively, and $\hat{I}_{\min}(t)$ and $\hat{I}_{\max}(t)$ are the lower and upper bounds for the true infected state $I(t)$, respectively, for all $t\geq 0$.}
\end{assumption}

\vspace{-1ex}
Moreover, we use $S^*(t)$, $I^*(t)$, $R^*(t)$ to represent the true states under the optimal control strategy $u^*(t)$, for all~$t\geq 0$, for the problem defined in~\eqref{eq:prob}. In order to propose a robust control strategy \baishe{with uncertainty}, we first explore the feasibility and control cost of the optimal control framework.
\vspace{-1.5ex}
\subsection{Feasibility and the Optimal \update{Control} Strategy}
\vspace{-2ex}
We first study the optimal control framework in \eqref{eq:prob} under \textit{accurate} model parameters and states. 
Let $t=0$ denote the very beginning
of an epidemic, and $t_p$ denote the time when the infection state reaches the peak value during the epidemic spreading process, i.e., $I(t_p)\geq I(t)$, $\forall t\geq 0$. The following lemma characterizes the peak value $I(t_p)$ 
in $\eqref{Eq: Con_Dynamics}$.
\vspace{-1ex}
\begin{lemma}
\label{lem:Ip}
Starting from $x(t_a)=[S(t_a)\quad I(t_a)\quad R(t_a)]^{\top}$ and $u(t_a)=u_{\fix}$ at time $t_a<t_p$, if the system in \eqref{Eq: Con_Dynamics} under the fixed control input $u(t)=u_{\fix}$ reaches a peak infection value $I(t_p)$, $u_{\fix}\in[0,\bar{u}]$ ,we have $I(t_p) = \rho (\ln\rho -1- \ln S(t_a))+S(t_a)+I(t_a)$, where $\rho=\frac{\gamma+u_{\fix}}{\beta}$.
\end{lemma}
\vspace{-4ex}
\begin{pf}
Consider \eqref{Eq: Con_Dynamics} for all $t\geq t_a$, dividing \eqref{eq:I_u} by  \eqref{eq:S_u} \bshee{gives} 
\[
\frac{dI(t)}{dS(t)}= \frac{\gamma+u(t)}{\beta S(t)}-1.
\]
Then, we integrate the equation with respect to $S(t)$ and apply the initial conditions $x(t_a)=[S(t_a)\quad I(t_a)\quad R(t_a)]^{\top}$ and $u(t_a)=u_{\fix}$. \baishe{Therefore,} by fixing $u(t)=u_{\fix}$, we obtain
\vspace{-2ex}
\begin{align*}
    &I(t) = \frac{\gamma+u_{\fix}}{\beta} \ln S(t) -S(t)\\
    &-\frac{\gamma+u_{\fix}}{\beta} \ln S(t_a)+S(t_a)+I(t_a), 
\vspace{-4ex}
\end{align*}
$\forall t\geq t_a$.
From 
\eqref{eq:I_u},
the infected population at $t_p$ satisfies 
\vspace{-1ex}
\[
\frac{dI(t_p)}{dt}=\beta S(t_p)I(t_p)-(\gamma+u_{\fix})I(t_p)=0,\]and $I(t_p)\neq0$. Hence, we have $S(t_p)=\frac{\gamma+u_{\fix}}{\beta}$ at $t_p$. 
By evaluating $I(t)$ at $t_p$ and substituting in $S(t_p)=\frac{\gamma+u_{\fix}}{\beta}=\rho$, we have that
$I(t_p) = \rho \ln\rho -\rho-\rho \ln S(t_a)+S(t_a)+I(t_a).$
We complete the proof.
\end{pf}
\vspace{-1.5ex}
Lemma \ref{lem:Ip}
calculates
the peak infection value $I(t_p)$ from any initial condition $x(t_a)$ under the fixed control input $u(t)=u_{\fix}$, for all $t\geq0$, before $t_p$, along with the transmission rate $\beta$ and the removal rate $\gamma$.
We can also generalize Lemma~\ref{lem:Ip} to compute the peak infection value $I(t_p)$ from any initial condition $x(t_a)$ under a different fixed control input $u(t_a)$, for all~$t\in [0, t_p)$.
Note that if $u_{\fix}=0$ for all $t\geq 0$, Lemma~\ref{lem:Ip} gives the peak infection value for the classic $SIR$ model.
\vspace{-1.5ex}
\begin{coro}
\label{Lem: cor}
Assume the closed-loop system in~\eqref{Eq: Con_Dynamics} starts from $x(t_a)=[S(t_a)\quad I(t_a)\quad R(t_a)]^{\top}$ and $u(t_a)=u_{\fix}$ at time $t_a$.
If there exists a time step $t_{p}$ such that~$I(t_{p})\geq I(t)$, $\forall t\geq t_a$, 
the peak infection value $I(t_{p})$ will increase as $\beta$ increases; decrease as $\gamma$ increases; and decrease as $u_{\fix}$ increases.
\end{coro}
\vspace{-4ex}
\begin{pf}
Consider $I(t_{p})$ as a function of $\rho$ in Lemma~\ref{lem:Ip}. Since $I(t_p)$ is the peak infection value during the epidemic spreading process, and $t_p>t_a$, then we have $\frac{dI(t)}{dt}>0$, for all $t\in [t_a, t_p)$. 
From \eqref{eq:I_u}, we have $\frac{\gamma+u_{\fix}}{\beta S(t)}< 1$, for all $t\in[t_a, t_{p})$,
\baishe{since the infected proportion will increase until reaching the peak infection value $t_p$.}
Define \baishe{the} function $g(\rho)=\rho \ln\rho-\rho-\rho \ln S(t_a)$, \baishe{where $\frac{\rho}{S(t)}\in  (0,1)$}, for all $t\in[t_a, t_{p})$. We obtain that the first derivative
$g'(\rho)=\ln\frac{\rho}{S(t_a)} 
<0$, since $\frac{\rho}{S(t_a)}\in (0,1)$. Therefore,
$g(\rho)$ is monotonically decreasing with respect to $\rho$, and thus $g(\rho)$ is monotonically decreasing with respect to $u_{\fix}$. Further, $I(t_p)$ is monotonically decreasing with respect to $\gamma$ and $u_{\fix}$, and monotonically increasing with respect to $\beta$, for all $t\in [t_a, t_p)$.
\end{pf}

\vspace{-1ex}
Corollary \ref{Lem: cor} implies that, 
under the same initial conditions, 
the peak infection value $I(t_p)$ will decrease with higher $\beta$ and/or lower $\gamma$. 
Further, 
Corollary~\ref{Lem: cor} states that increasing the \update{fixed isolation rate} $u_{\fix}$ will lower the peak infection value. \update{Hence, if $I(t_p)\leq \bar{I}$, with a fixed isolation rate of $u(t)=u_{\fix}=0$,  for all~$t\geq 0$,
the optimal control strategy will be $u(t)=0$, for all~$t\geq 0$.}
\vspace{-2ex}
\begin{coro}[Optimal \update{Control} Strategy 1]
\label{lem:opt1}
The optimal \update{control} strategy for the problem in \eqref{eq:prob} is $u^*(t)=\update{0}$, for all~$t\geq 0$, if $I^*(t_p) = \rho (\ln\rho -1- \ln S^*(0))+S^*(0)+I^*(0)\leq \bar{I}$.
\end{coro}
\vspace{-1ex}
Corollary~\ref{lem:opt1} is a direct result from Lemma \ref{lem:Ip} and Corollary~\ref{Lem: cor}. For the optimal control problem in \eqref{eq:prob}, if there is no risk for the infection state to exceed the infection threshold $\bar{I}$ \update{without any isolation strategy}, 
maintaining the \update{isolation} rate at zero is the best way to reduce the \update{isolation} cost. For the control framework in~\eqref{eq:prob}, 
we consider the case \bshee{when} $I(t_p)> \bar{I}$ under \update{$u(t)=0$}, for all~$t\geq 0$, 
and develop the following theorem to study the feasibility of the control problem in \eqref{eq:prob}.
\vspace{-1.5ex}
\begin{thm}
\label{Thm:feasibility}
Starting from $t_a\geq 0$, if there exists $t_b\geq t_a$ such that $I(t_b)=\bar{I}$ for the first time, then
the control framework in \eqref{eq:prob} is feasible if and only if there exists a control input~$u(t_b)\in(\update{0}, \bar{u}]$ such that $u(t_b)=\beta S(t_b)-\gamma$.
\end{thm}
\vspace{-4ex}
\begin{pf}
Consider the system in~\eqref{Eq: Con_Dynamics} before reaching $t_b$, we have $I(t)<\bar{I}$, for all $t\in [0, t_b)$. 
Hence, the system is feasible for all~$t\in [0, t_b)$. Then we study the system starting from $t_b$.\\
$\Longleftarrow$: Under the condition that there exists $t_b\geq t_a$ such that $I(t_b)=\bar{I}$ for the first time, if there exists $ u(t_b)\in(\update{0}, \bar{u}]$ such that $u(t_b)=\beta S(t_b)-\gamma$, from~\eqref{eq:I_u},
we have $\frac{dI(t_b)}{dt}=0$. 
Furthermore, since $S(t)$ is strictly monotonically decreasing unless $S(t)=0$ and/or $I(t)=0$  for all~$t\geq0$, we can always find a $u(t)\in [u(t_b), \bar{u}]$, such that
$u(t)\geq \beta S(t)-\gamma$ for all~$t\geq t_b$. 
From \eqref{eq:I_u}, there  always exists a $u(t)\in  [u(t_b), \bar{u}]$ such that $\frac{dI(t)}{dt}\leq 0$, for all $t\geq t_b$, which guarantees
$I(t)\leq \bar{I}$ for all $t\geq t_b$. Therefore, the control framework in \eqref{eq:prob} is feasible.\\
$\Longrightarrow$: Starting from $t_a\geq 0$, there exists $t_b>t_a$ such that $I(t_b)=\bar{I}$ for the first time. If
the system is feasible, $I(t)$ must stop increasing at $t_b$. Hence, $\frac{dI(t_b)}{dt}\leq 0$ indicates that there must exist $u(t_b)\in (\update{0},\bar{u}]$ such that $\frac{dI(t_b)}{dt}=\beta S(t_b)\bar{I}-(\gamma+u(t_b))\bar{I}=0$. Therefore, we have
$u(t_b)=\beta S(t_b)-\gamma$, which completes the proof.
\end{pf}
\vspace{-1ex}
In this work, we study the case that satisfies Theorem~\ref{Thm:feasibility}:
the upper bound on the \update{isolation} rate $\bar{u}$ is \bshee{sufficiently large} such that we can always find a $u(t_b)\in(\update{0}, \bar{u}]$, to satisfy $u(t_b)= \beta S(t_b)-\gamma$. Under such condition, the optimal \update{control} strategy is given by the following proposition, where $a^*\in \{S^*,I^*,R^*,t^*_b,t^*_h\}$ represents the true state or the time step of the system in \eqref{Eq: Con_Dynamics}
under the optimal control strategy $u^*(t)$.
Note that $t^*_b$ is the time step when $I^*(t)$, $t\geq 0$, reaches $\bar{I}$ under the optimal \update{control} strategy $u^*(t)$ for the first time. In addition, $t^*_h$ is the time step  when the epidemic reaches \baishe{our defined herd immunity time} under the optimal \update{control} strategy $u^*(t)$ for the first time, i.e., $\frac{dI(t^*_h)}{dt}=(\beta S(t^*_h)-\gamma)I(t^*_h)=0$.
\baishe{Herd immunity is a concept in epidemiology that describes a situation where a large proportion of a population becomes immune to a contagious disease, either through vaccination or previous exposure to the disease~\cite{fine1993herd}. When a significant portion of the population is immune, the spread of the disease is significantly slowed or stopped because there are fewer susceptible individuals for the pathogen to infect. In this work, we define the herd immunity time step $t^*_h$ as the moment when the susceptible proportion is sufficiently small such that the number of infected cases starts to decrease \update{without any isolation, i.e., $u(t)=0$.}} Furthermore, we have $\frac{dI(t^*_h)}{dt}\leq 0$, for all $u(t)\in[0, \bar{u}]$, and $t\geq t^*_h$. Under these settings, we introduce the following optimal control strategy.
\vspace{-1.5ex}
\begin{proposition}[Optimal \update{Control} Strategy 2]
\label{Prop:policy}
\cite[Theorem 1]{acemoglu2021optimal}
The optimal \update{control} strategy for the problem in \eqref{eq:prob} can be cast into three stages:
\vspace{-1.5ex}
\begin{enumerate}
    \item At the early stage of the epidemic,  when $I^*(t)<\bar{I}$, for all $t\in[0, t^*_b)$, $u^*(t)=\update{0}$;
    \item During the outbreak, starting from $I^*(t^*_b)=\bar{I}$, for all $t\in[t^*_b, t^*_h)$, $u^*(t)=\beta S^*(t)-\gamma$;
    \item When the epidemic reaches herd immunity at $t^*_h$, i.e., $\beta S^*(t^*_h)=\gamma$, for all $t\geq t^*_h$, $u(t)=\update{0}$.
\end{enumerate}
\end{proposition}

\vspace{-1.5ex}
The proof of Proposition \ref{Prop:policy} is the same as the proof of \cite[Theorem 1]{acemoglu2021optimal}.
Proposition \ref{Prop:policy} separates the \update{isolation process} into three stages via considering the first time when the infection state reaches $\bar{I}$, i.e., $t^*_b$, and the herd immunity time step $t^*_h$ as the switching time steps. 
In the following section, we aim to explore a robust \update{isolation} strategy under the guidance of the optimal \update{isolation} strategy in Proposition \ref{Prop:policy}, \baike{with parameter error bounds in~\eqref{eq:error_bounds} 
and the state uncertainties given by Assumption~\ref{Assum_bounds}.}
\vspace{-2ex}
\subsection{Robust \update{Control} Strategy}
\begin{figure*}
    \centering
\includegraphics[ trim = 2cm 1.1cm 2cm 0.5cm, clip, width=0.9\textwidth]{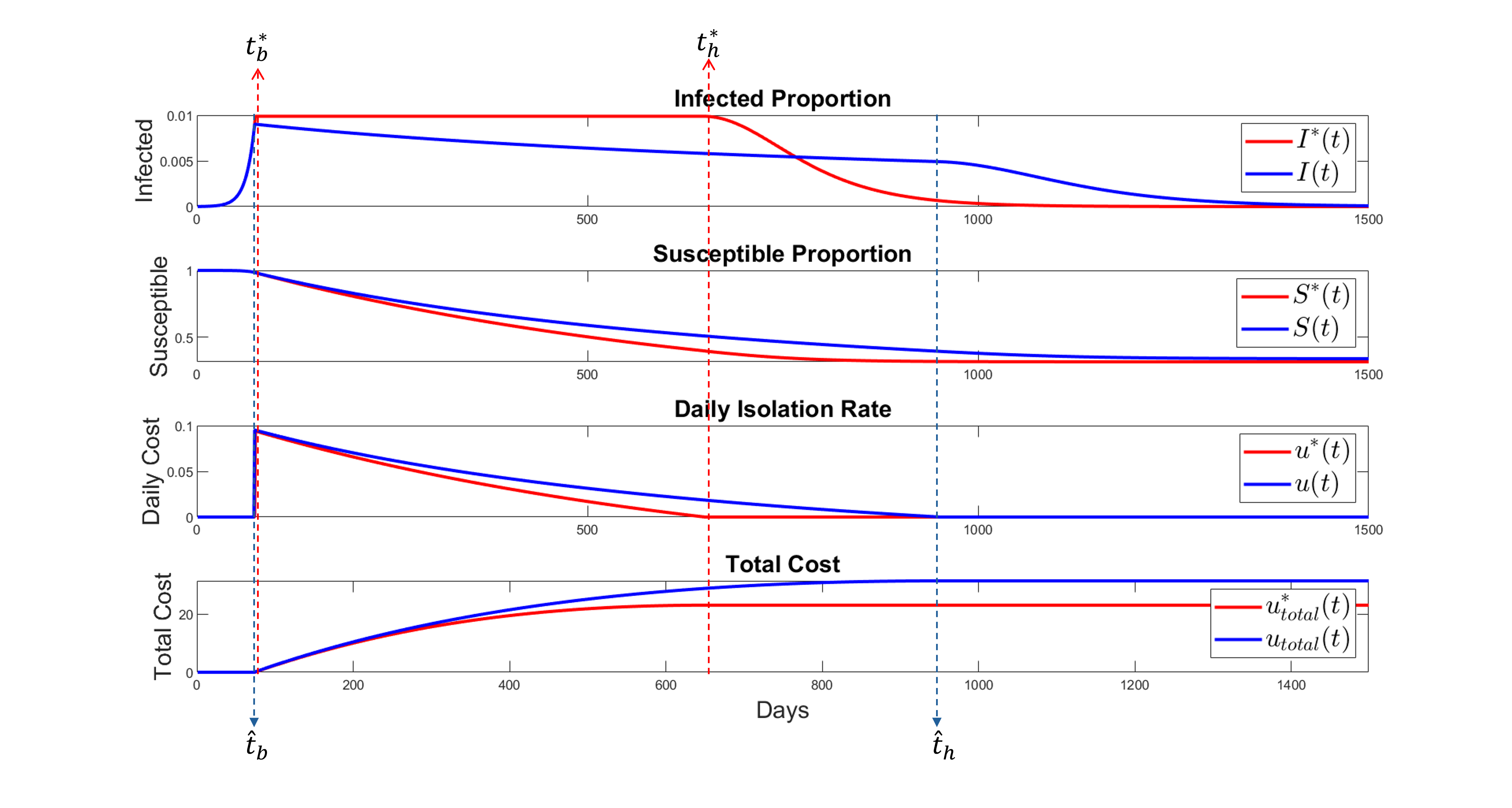}
    \caption{Comparison of Theorem \ref{lem:inaccu} with the Optimal \update{Control} Strategy. \baishe{The figure illustrates the difference between the optimal \update{control} strategy and the corresponding spreading process and cost, marked with red lines, and the robust \update{control} strategy and its corresponding spreading process and cost, marked with blue lines. The optimal \update{control} strategy begins to raise the \update{isolation} rate when the infected proportion reaches the infection threshold at time step $t^*_b$. To maintain the number of infected cases at the infection threshold $0.1$, during the outbreak, the optimal \update{control} strategy adjusts the \update{isolation} rate between $t^*_b$ and $t^*_h$. The \update{isolation} rate is then adjusted back to the lowest level after $t^*_h$. In contrast, the robust \update{control} strategy adjusts the \update{isolation} rate ahead of $t^*_b$, captured by $\hat{t}_b$. This strategy also maintains a higher \update{isolation} rate no later than $t^*_h$. Furthermore, the total daily \update{isolation} rate of the optimal \update{control} strategy is upper-bounded by the daily \update{isolation} rate of the robust \update{control} strategy. Therefore, the total cost of the
    optimal \update{control} strategy
    is upper-bounded by the total cost of the robust \update{control} strategy.}}
    \label{fig:example}
\label{-1ex}
\end{figure*}

\vspace{-2ex}
In this section, we propose a robust \update{control} strategy for the problem in~\eqref{eq:prob}. \bshee{Recall that we define $\hat S(t)$, $\hat{I}(t)$, $\hat{R}(t)$, for all $t\geq 0$ as the measured states. We use $\hat{t}_b$ to denote the time step when the overestimated state $\hat I_{\max} (t)$ reaches the infection threshold $\bar{I}$ for the first time. In addition, we use $\hat{t}_h$ to represent the time step  when
\vspace{-1.5ex}
\begin{equation}
\label{eq:herd}
    \hat{\beta}_{\max} \hat{S}_{\max}(\hat{t}_h)=\hat{\gamma}_{\min},
\vspace{-1.5ex}
\end{equation}
for the first time, i.e., the computed herd immunity time step by overestimating the epidemic states and spreading parameters. We use $\hat{u}(t)$, for all $t\geq 0$, to represent the generated \update{isolation} strategy by leveraging the overestimated epidemic spreading process and the corresponding computed time steps $\hat t_h$ and $\hat t_b$.}
\vspace{-2ex}
\begin{definition}[Robust \update{Control} Strategy]
\label{Def: T-U}
The \update{control} strategy for the problem in \eqref{eq:prob} follows the rules:
\vspace{-1.5ex}
\label{def:optimal_p}
\begin{enumerate}
     \item At the early stage of the epidemic, \bshee{when the overestimated infection state is smaller than the infection threshold $\bar{I}$, the \update{isolation rate} is given by $\hat u(t)=0$, for all $t\in [0, \hat{t}_b)$; 
    \item From the time step $\hat{t}_b$ to the computed herd immunity time step $\hat{t}_h$,
    the \update{isolation rate} is given by 
    $\hat u(t)=\hat{\beta}_{\max}\hat{S}_{\max}(t)$, for all~$t\in[\hat{t}_b, \hat{t}_h)$;}
    \item Starting from the computed herd immunity time step~$\hat{t}_h$, the \update{isolation rate} is given by  $\hat u(t)=0$, for all $t\geq \hat{t}_h$. 
\end{enumerate}
\end{definition}
\vspace{-1.5ex}
Definition \ref{def:optimal_p} modifies the optimal \update{control} strategy in Proposition~\ref{Prop:policy} by proposing a \update{isolation} policy under the given bounds of estimated parameters and states. Definition \ref{def:optimal_p} implies that without accurate model parameters and states, \baike{if we know the bounds of the parameters and states,} the \update{control} strategy will always assume the worst-case 
scenario at any given time step to generate the \update{isolation} policy, i.e., to overestimate the seriousness of the epidemic. Note that Lemma~\ref{lem:Ip}, Corollary~\ref{Lem: cor}, and Definition~\ref{Def: T-U}  partially solve Problem~\ref{Prob-2} by proposing a robust \update{control} strategy.
\vspace{-1.5ex}
\begin{remark}
    The optimal \update{control} strategy in  Proposition~\ref{Prop:policy} is not robust against parameter and state uncertainties. For example, during stage~2 in Proposition~\ref{Prop:policy}, the optimal \update{isolation} rate is given by $u^*(t) = \beta S^*(t) -\gamma$.
    However, in \baishe{a} real-world epidemic spreading process, it is challenging to access accurate model parameters and states.
    Parameter estimation errors in $\beta$ and $\gamma$ may result in an insufficient \update{isolation} rate, potentially leading to outbreaks~\cite{morris2021optimal}.
    We will show that
    one advantage of the proposed robust \update{control} strategy in Definition~\ref{def:optimal_p} is to ensure the feasibility of the control design.
\end{remark}
\vspace{-1.5ex}
\begin{remark}
One way to obtain the error bounds of the model parameters is given by Theorem~\ref{Error_bound} and Corollary~\ref{coro: range}. 
For instance, at the beginning of a pandemic, we can sample the susceptible and infected states to estimate the spreading parameters and error bounds, since the robust control strategy in Proposition~\ref{Prop:policy} maintains the \update{isolation} rate at zero during the early state of the pandemic.
Then, the estimated error bounds can be leveraged for policy-making during stages 2 and 3 in Definition~\ref{Def: T-U}.  
\end{remark}
\vspace{-1ex}
We discuss the feasibility of the system in~\eqref{Eq: Con_Dynamics} under the robust \update{control} strategy in Definition~\ref{def:optimal_p} by  studying the situation where we know
$\hat{\beta}_{\max}$, $\hat{\gamma}_{\min}$, 
and $\hat S(t) \in[S(t),\hat{S}_{\max}(t)]$, $\hat I(t) \in[I(t),\hat{I}_{\max}(t)]$, for all $t\geq 0$. 
We 
plot
both trajectories of the system under the optimal \update{control} strategy $u^*(t)$ and the strategy $\hat u(t)$ from Definition~\ref{def:optimal_p} in Figure~\ref{fig:example}, in order to better explain  $t^*_b$, $t^*_h$, $\hat t_b$, and $\hat t_h$. \bks{ Figure~\ref{fig:example} compares the behavior of the epidemic under the \update{control} strategy in Definition~\ref{def:optimal_p} when overestimating the spreading parameters and states, with the behavior of the epidemic under the optimal \update{control} strategy in Proposition~\ref{Prop:policy} when the true spreading parameters and states are known. Consider an epidemic spreading process with $\beta=0.16$ and $\gamma=0.063$. The infection threshold is set as $\bar{I}=0.01$. The lower and upper bounds on the isolation rate are \update{zero} and $\bar{u}=0.2$, respectively. We use $S^*(t)$, $I^*(t)$, and $R^*(t)$, $t\geq0$, to represent the states generated by $u^*(t)$ following
the Optimal \update{Control} Strategy~1 in Proposition~\ref{Prop:policy}. We use $S(t)$,
$I(t)$, and $R(t)$, $t\geq0$, to denote the true states  generated by $\hat u(t)$, when implementing 
the \update{control} strategy given in Definition~\ref{def:optimal_p}
and leveraging the overestimated spreading parameters $\hat{\beta}(t)=\hat{\beta}_{\max}=1.05\beta$ and  $\hat{\gamma}(t)=\hat{\gamma}_{\min}=0.95\gamma$,
and noisy measurements $\hat{S}(t)$ and $\hat{I}(t)$, for all $t \geq 0$. \baishe{The variance of the noise for the susceptible state and infected state at time step~$t$ are given by~$S(t)/100$ and~$I(t)/100$, respectively. We use relatively small noise here for the purpose of illustrating the idea.}} 
From Definition~\ref{def:optimal_p}, we \bshee{will leverage Figure~\ref{fig:example} to illustrate} the following result.
\vspace{-1.5ex}
\begin{thm}
\label{lem:inaccu}
When 
$\hat S(t) \in[S(t) ,\hat{S}_{\max}(t)]$, $\hat I(t) \in[I(t),\hat{I}_{\max}(t)]$,  for all $t\geq 0$, the system in \eqref{Eq: Con_Dynamics} under the control strategy $\hat u(t)$ generated by leveraging  $\hat{\beta}_{\max}$, $\hat{\gamma}_{\min}$, $\hat{S}(t)$, $\hat I(t)$, for all $t\geq0$,
from Definition~\ref{def:optimal_p} is feasible. The control strategy satisfies $\hat u(t)\geq u^*(t)$, for all $t\geq0$.
\end{thm}
\vspace{-3.5ex}
\begin{pf}
We compare $u^*(t)$ and $\hat{u}(t)$ by considering $t\in [0, \hat t_b)\cup [\hat t_b, t^*_b]\cup [t^*_b, t^*_h]\cup(t^*_h, \hat t_h] \cup (\hat t_h, +\infty)$, where the chronological order will be demonstrated within the context.
First, we show the system in \eqref{Eq: Con_Dynamics} under the \update{control} policy $\hat u(t)$, for all $t\geq 0$, is feasible. We analyze the \update{control} strategy by considering three main stages. Recall that the control framework first switches its \update{control} policy when $\hat I(\hat t_b)=\bar{I}$ ($\hat{t}_b$ is the first time when $\hat{I}(t)$ reaches $\bar{I}$, as shown in the \bshee{top plot of} Figure~\ref{fig:example}). 
Since $\hat I(t)\geq I(t)$, for all $t\geq 0$, we have
\vspace{-2ex}
\begin{equation*}
 I(\hat{t}_b)=I(\hat{t}_b)\leq \hat I(\hat{t}_b)=\bar{I}.
\vspace{-2ex}
\end{equation*}
Hence, compared to using the optimal \update{control} policy $u^*(t)$, for all $t\in [0, t^*_b]$, the system, by leveraging larger estimated infection states, will start to raise the \update{isolation rate} away from \update{zero} earlier, i.e., at $\hat{t}_b$. 
Hence, we have $\hat{t}_b\leq{t}^*_b$, as illustrated in Figure \ref{fig:example}. In addition, 
at the early stage of the epidemic, when $\hat I(t)<\bar{I}$, for all~$t\in[0, \hat t_b)$, we have $\hat u(t)=u^*(t)=0$. 

\vspace{-1.5ex}
Then, we consider the time step when $\hat I(\hat t_b)=\bar{I}$. 
From
Definition \ref{def:optimal_p}, we have
\vspace{-1.5ex}
\begin{equation*}
(\beta S(t)-(\gamma+\hat u(t))\leq (\hat{\beta}_{\max} \hat{S}(t)-(\hat{\gamma}_{\min}+\hat u(t)) =0.   
\vspace{-1.5ex}
\end{equation*}
Thus $\frac{dI(t)}{dt}\leq 0$, for all $t\in [\hat t_b, \hat{t}_h]$, where
$\hat{t}_h$ is the computed herd immunity time step under the condition that $\hat{S}(\hat{t}_h)\hat{\beta}_{\max}-\hat{\gamma}_{\min}=0$ (shown in Figure \ref{fig:example}). Hence, the infection state $I(t)$ is non-increasing under $\hat u(t)$, and  $I(t)\leq \bar I$, for all $t\in [\hat {t}_b, \hat{t}_h]$. Lastly, after reaching the computed herd immunity time step $\hat t_h$, from Definition~\ref{def:optimal_p}, we have $\hat u(t)=0$ and 
\vspace{-2ex}
\begin{equation*}
\vspace{-2ex}
(\beta S(t)-\gamma)\leq (\hat{\beta}_{\max}\hat{S}(t)-\hat{\gamma}_{\min})\leq 0, 
\end{equation*} 
for all $t\geq \hat{t}_h$. 
Therefore,
$I(t)$, for all $t\geq \hat{t}_h$ will monotonically decrease, and thus cannot reach $\bar{I}$ again. The trajectories of the optimal states under $u^*(t)$ and the true states under the \update{control} strategy $\hat{u}(t)$, for all $t\geq 0$, are shown in Figure \ref{fig:example}. 
In summary, starting from $t=0$, $I(t)$ cannot exceed $\bar{I}$ under the given control policy $\hat u(t)$, for all $t\geq 0$, which completes the proof of the feasibility. 

\vspace{-1ex}
Now we compare the \update{isolation} rate $\hat u(t)$ from Definition~\ref{Def: T-U} and the optimal \update{isolation} rate $u^*(t)$ from Proposition~\ref{Prop:policy}.  Recall at the early stage of the epidemic, when $\hat I(t)<\bar{I}$, for all $t\in[0, \hat{t}_b)$, $\hat u(t)=u^*(t)=0$.
Starting from $\hat t_b$, we have that
\vspace{-2.5ex}
\begin{equation*}
\hat u(t)=\hat S(t)\hat\beta_{\max}-\hat\gamma_{\min}\geq 0=u^*(t), 
\vspace{-2ex}
\end{equation*}
for all $t\in[\hat{t}_b, t^*_b]$. Note that $\hat u(t)$ is not the optimal control strategy (but a strategy that ensures the system is feasible) for the problem defined in \eqref{eq:prob}. Moreover, \cite[Lemma~8]{acemoglu2021optimal} shows that, among all the feasible frameworks, the system in \eqref{Eq: Con_Dynamics} reaches the herd immunity time step $t^*_h$ the fastest, under the optimal \update{control} strategy $u^*(t)$. Hence, we have $\hat{t}_h\geq t_h\geq t^*_h$. Recall that $\hat S(t)$ and $S(t)$ for all $t\geq 0$ are the estimated susceptible state and the corresponding true state under the control policy from Theorem \ref{lem:inaccu}, respectively.
In addition, $t_h$ and $\hat{t}_h$  are the time steps when $S(t_h)\beta-\gamma=0$ and $\hat{S}(\hat{t}_h)\beta_{\max}-\gamma_{\min}=0$ under the control policy $\hat u(t)$, respectively. The inequality $\hat{t}_h\geq t_h$ implies that when $S(t_h)\beta-\gamma=0$, the estimated parameters and states still satisfy 
\vspace{-1.5ex}
\begin{equation*}
\hat S(t_h)\hat\beta_{\max}-\hat\gamma_{\min}\geq0.
\vspace{-1.5ex}
\end{equation*}
Thus, 
compared to $t_h$, it will take longer for the system to reach the estimated herd immunity time step $\hat t_h$. 
From Proposition \ref{Prop:policy} and Definition~\ref{def:optimal_p},  $u^*(t)=0$, for all $t\geq t^*_h$, and 
\vspace{-3ex}
\begin{equation*}
\hat u(t)=\hat{S}(t)\hat{\beta}_{\max}-\hat{\gamma}_{\min}\geq 0, 
\vspace{-1.5ex}
\end{equation*}
for all $t\in [t^*_h, \hat{t}_h]$. In addition, we have $\hat u(t)=0$, for all $t\geq \hat{t}_h$, which  leads to $0=u^*(t)\leq \hat u(t)$, for all $t\geq t^*_h$. 

\vspace{-1ex}
Lastly, we analyze both \update{control} policies when $t\in [t^*_b, t^*_h]$. Following the discussion from the feasibility and the fact that the optimal control strategy $u^*(t)$ maintains $I^*(t)=\bar{I}$ for all $t\in [t^*_b, t^*_h]$, we have 
$I(t)\leq\bar{I}=I^*(t)$,  for all 
$t\in[t^*_b, t^*_h]$. 
Hence, 
\baishe{by dividing both sides of the integration of~\eqref{eq:S_u} by $S(t)$, we obtain the following equation:}
\vspace{-2.5ex}
\begin{equation*}
 log (S(t))=log (S(t^*_b))-\int_{t^*_b}^{t}(\beta I(\tau)) d\tau.
\vspace{-2ex}
\end{equation*}
\baishe{If} $I(t)\leq\bar{I}=I^*(t)$,  for all $t\in [t^*_b, t^*_h]$, then 
\vspace{-2ex}
\begin{equation*}
\hat S(t)\geq S(t)\geq S^*(t),  
\vspace{-2ex}
\end{equation*}
for all $t\in [t^*_b, t^*_h]$ (note that $S(t^*_b)\geq S^*(t^*_b)$).
From 
the fact that $\hat S(t)\geq S^*(t)$, for all $t\in [t^*_b, t^*_h]$, and  $\hat{\beta}_{\max}\geq\beta$, $\hat{\gamma}_{\min}\leq\gamma$, we have $\hat u(t)\geq u^*(t)$, for all $t \in [t^*_b,t^*_h]$. \baishe{This completes} the proof.
\end{pf}

\vspace{-1ex}
Theorem \ref{lem:inaccu} explores the case where the estimated upper and lower bounds on the parameters $\beta_{\max}$ and $\gamma_{\min}$ and states are known. 
\baishe{The theorem} implies that \update{the isolation rates} $\hat u(t)=u^*(t)= 0$, for all~$t\in [0, \hat{t}_b)\cup [\hat{t}_h, +\infty)$. 
In addition, compared to $u^*(t)$, the proposed \update{control} policy $\hat{u}(t)$ from Theorem~\ref{lem:inaccu} starts to raise the \update{isolation} rate from \update{zero} earlier (at $\hat{t}_{b}$), and switches back to \update{zero} later (at $\hat{t}_{h}$). Hence, Theorem~\ref{lem:inaccu} demonstrates that the proposed \update{robust control} strategy in Definition~\ref{Def: T-U} sacrifices optimality but guarantees the feasibility of the optimal control problem in~\eqref{eq:prob}. By considering the worst-case scenario, the robust \update{control} strategy from Definition~\ref{Def: T-U} can always ensure feasibility if the true parameters and states satisfies $\beta\in [\hat\beta_{\min},\hat\beta_{\max}]$, $\gamma\in [\hat\gamma_{\min},\hat\gamma_{\max}]$, $S(t)\in [S_{\min}(t),S_{\max}(t)]$, $I(t)\in [I_{\min}(t),I_{\max}(t)]$. 
Further, 
Theorem~\ref{lem:inaccu} infers that the cost of $\hat u(t)$ depends on the parameter estimation error bounds $\hat\beta_{\max}$ and $\hat\gamma_{\min}$, and the state measurement bounds $\hat{S}_{\max}(t)$ and $\hat{I}_{\max}(t)$. 

\vspace{-1.5ex}
\begin{remark}   
Recall that Theorem~\ref{Error_bound} and Corollary~\ref{coro: range} demonstrate that the estimation error bounds on $\hat{\beta}$ and $\hat{\gamma}$ are determined by the sample step size $h$ and the state measurement error $v_{\max}$. Smaller sample step size $h$ will lower the estimation error bound (supposing that the measurement error is small). 
The proof of Theorem~\ref{lem:inaccu} further implies that the increment on error bounds of the parameters and states has an impact on the robust control algorithm in Definition~\ref{Def: T-U}. 
First, the higher measurement error on $\hat I(t)$ will let the robust \update{control} strategy switch from \update{zero} to the rate in Step 2) earlier.  In addition, in Step 2), higher error bounds on estimated parameters and states will result in more \update{isolation} resources. 
Further, 
it takes longer to switch the \update{isolation} rate in Step~2) back to \update{zero} in Step~3) under higher error bounds on the estimated parameters and $\hat{S}(t)$, since the computed herd immunity time $\hat{t}_h$ is also related to the bounds on parameters and states, as indicated in~\eqref{eq:herd}.
In summary, as outlined in the proof of Theorem~\ref{lem:inaccu}, as the parameter estimation error and state measurement error become larger, 
the robust \update{control} strategy in Definition~\ref{Def: T-U} will be farther away from the optimal \update{control} strategy. 
\end{remark}
\vspace{-2ex}
\subsection{Additional \update{Control} Cost}
\vspace{-2.5ex}
In this section, we further quantify the additional cost of implementing our proposed robust \update{control} strategy\baishe{, compared to the optimal strategy from Proposition~\ref{Prob-1}.}
\vspace{-1.8ex}
\begin{lemma}
\label{lem:cost}
The overall cost by leveraging $\hat{\beta}_{\max}\geq\beta$, $\hat{\gamma}_{\min}\leq\gamma$, $\hat S(t) \in[S(t),\hat{S}_{\max}(t)]$, $\hat I(t) \in[ I(t),\hat{I}_{\max}(t)]$ for all $t\geq 0$, is higher than the optimal cost by
\vspace{-2ex}
\begin{equation*}
\mathcal{C}=\int_{\hat{t}_b}^{\hat{t}_h}(\beta (S(t)-S^*(t))) dt 
-log(I(\hat{t}_h))+log(I^*(\hat{t}_h)).
\vspace{-5ex}
\end{equation*}
\end{lemma}
\begin{pf}
From the Proof of Theorem~\ref{lem:inaccu}, $\hat u(t) = u^*(t)$ when $t\in [0,\hat t_b) \cup (\hat t_h,+\infty]$. Hence, when comparing 
the additional cost, we study the difference from $\hat{t}_b$ to $\hat{t}_h$. 
From \eqref{eq:I_u}, we have
\vspace{-2ex}
\begin{equation}
\label{eq: testing_rate}
  u(t)=-\frac{1}{I(t)}\frac{dI(t)}{dt}+\beta S(t) -\gamma. 
\vspace{-2ex}  
\end{equation}
By integrating this equation, and comparing $\hat{u}(t)$ and $u^*(t)$ within the range of $[\hat t_b,\hat t_h ]$,
we have 
\vspace{-2ex}
\begin{align*}
\mathcal{C}=\int_{\hat{t}_b}^{\hat{t}_h} (\hat u(t)-u^*(t))dt\ \ \ \ \ \ \  \ \ \ \ \ \ \ \ \ \ \ \ \ \ \ \ \ \ \ \  \ \ \ \ \ \ \ \ \ \ \\  =
\int_{\hat{t}_b}^{\hat{t}_h}(\beta (S(t)-S^*(t))) dt 
-log(I(\hat{t}_h))+log(I^*(\hat{t}_h)),
\vspace{-4ex}
\end{align*}
where $log(I(\hat{t}_b))=log(I^*(\hat{t}_b))$ is used.
\end{pf}
\vspace{-1ex}
Lemma~\ref{lem:cost} states that the difference between $\hat u(t)$ and $u^*(t)$ is captured by the difference between the susceptible states $S(t)$ and $S^*(t)$, and the infection states when the systems reach the computed herd immunity time step $\hat{t}_h$. 
Recall that Theorem~\ref{lem:inaccu}
studies the system's feasibility under the robust \update{control} strategy. Following Theorem~\ref{lem:inaccu}, 
Lemma~\ref{lem:cost} captures the additional cost $\mathcal{C}$ of the robust control strategy through the susceptible and infected states. In order to further build the connection between the additional cost $\mathcal{C}$ and the error bounds on the model parameters $\hat\beta_{\max}$, $\hat\gamma_{\min}$, 
and measured states $\hat{S}_{\max}(t)$ and $\hat{I}_{\max}(t)$,
we derive the following theorem.
\vspace{-1.5ex}
\begin{thm}
\label{thm: cost}
The overall cost by leveraging $\hat{\beta}_{\max}\geq\beta$, $\hat{\gamma}_{\min}\leq\gamma$, $\hat S(t) \in[S(t),\hat{S}_{\max}(t)]$, $\hat I(t) \in[\hat I(t),\hat{I}_{\max}(t)]$ for all $t\geq 0$, is higher than the optimal cost by
\vspace{-1ex}
\begin{align}
\mathcal{C}&=  \nonumber 
-\hat{\gamma}_{\min}(t^*_b-\hat{t}_b+\hat{t}_h-t^*_h)+ (-\hat{\gamma}_{\min}+\gamma)(t^*_h-t^*_b) \\ \nonumber 
& \ \ \ \ +\hat{\beta}_{\max} (\int_{\hat{t}_b}^{t^*_b}\hat{S}_{\max}(t)dt+\int_{t^*_h}^{\hat{t}_h}\hat{S}_{\max}(t)dt)\\ 
& \ \ \ \ + \int_{t^*_b}^{t^*_h}(\hat{S}_{\max}(t)\hat{\beta}_{\max} -S^*(t)\beta )dt.
\label{eq:add_cost}
\vspace{-5ex}
\end{align}
The optimal cost $\mathcal{C}$ further satisfies
\vspace{-2ex}
\begin{align*}
\vspace{-2ex}
\mathcal{C}\leq \bar{\mathcal{C}}&= (\hat{S}_{\max}(\hat{t}_b)\hat{\beta}_{\max}-\hat{\gamma}_{\min})(t^*_b-\hat{t}_b+\hat{t}_h-t^*_h) \\ \nonumber
&+ (\hat{S}_{\max}(\hat{t}_b)\hat{\beta}_{\max}-\hat{\gamma}_{\min}-S^*(t^*_h)\beta+\gamma) (t^*_h-t^*_b).
\end{align*}
\end{thm}
\vspace{-6ex}
\begin{pf}
From Definition~\ref{Def: T-U}, the robust \update{control} policy~$\hat{u}(t)$ during $[\hat{t}_b, \hat{t}_h]$ is given by $\hat{S}_{\max}(t)\hat{\beta}_{\max}-\hat{\gamma}_{\min}$. Hence,
the difference between $\hat u(t)$ and $u^*(t)$ is
\vspace{-2ex}
\begin{align*}
\vspace{-6ex}
\mathcal{C}&=\int_{\hat{t}_b}^{\hat{t}_h} (\hat u(t)-u^*(t))dt \\ \nonumber
&=\int_{\hat{t}_b}^{\hat{t}_h}(\hat{S}_{\max}(t)\hat{\beta}_{\max}-\hat{\gamma}_{\min})dt 
-\int_{t^*_b}^{t^*_h}(S^*(t)\beta-\gamma)dt 
\\ \nonumber
&= \int_{\hat{t}_b}^{t^*_b}(\hat{S}_{\max}(t)\hat{\beta}_{\max}-\hat{\gamma}_{\min})dt\\ \nonumber
&\ \ \ \ + \int_{t^*_b}^{t^*_h}(\hat{S}_{\max}(t)\hat{\beta}_{\max}-\hat{\gamma}_{\min}-(S^*(t)\beta-\gamma)  )dt \\ \nonumber
& \ \ \ \ + \int_{t^*_h}^{\hat{t}_h}(\hat{S}_{\max}(t)\hat{\beta}_{\max}-\hat{\gamma}_{\min})dt.
\vspace{-4ex}
\end{align*}
By reorganizing the equation and computing the integration, we can obtain first equation in the theorem.

\vspace{-1ex}
To derive the upper bound on the additional cost $\mathcal{C}$, we upper bound the estimated state $\hat{S}_{\max}(t)$ by $\hat{S}_{\max}(\hat{t}_b)$ in~\eqref{eq:add_cost}, and lower bound the true state $S^*(t)$ by $S^*(t^*_h)$, for all $t\in[t^*_b, t^*_h]$. 
Note that $\hat{t}_b$ is the moment when the computed infected proportion $\hat{I}_{\max}(t)$ reaches the infection threshold $\bar{I}$. For an $SIR$ model, the susceptible proportion is monotonically non-increasing. Thus, we have $\hat{S}_{\max}(t)\leq \hat{S}_{\max}(\hat{t}_b)$, 
for all $t\geq \hat{t}_b$, and $S^*(t)\geq S^*(t^*_h)$ for all $t\in[t^*_b, t^*_h]$.
By replacing $\hat{S}_{\max}(t)$ with $\hat{S}_{\max}(\hat{t}_b)$, and $S^*(t)$ by $S^*(t^*_h)$
in~\eqref{eq:add_cost}, we can obtain $\bar{\mathcal{C}}$. 
\end{pf}
\vspace{-1.5ex}
Theorem~\ref{thm: cost} characterizes the additional cost of the robust control
strategy through an explicit form. From Theorem~\ref{thm: cost}, 
the additional cost $\mathcal{C}$ is not only determined by the error bounds from the estimated parameters $\hat{\beta}_{\max}$ and $\hat{\gamma}_{\min}$, but also relies on the estimated states $\hat{S}_{\max}(t)$ and the computed switching time $\hat{t}_b$ and $\hat{t}_h$. 
Further, Theorem~\ref{thm: cost} 
develops the upper bound $\bar{\mathcal{C}}$ on the additional cost $\mathcal{C}$, which further confirms that 
the bounds on the additional cost
between $\hat u(t)$ and $u^*(t)$ are captured by 1) the time $\hat{t}_b$ and $\hat{t}_h$; 2) the estimated states $\hat{S}_{\max}(\hat{t}_b)$ and the optimal state $S^*(t^*_h)$; 3) the parameter bounds $\beta_{\max}$
and $\gamma_{\min}$. Note that $t^*_h$ and $t^*_b$ can be considered as constants, which are determined by the spreading behavior under the optimal control strategy $u^*(t)$. In addition, Theorem~\ref{thm: cost} shows that, when the estimated parameters equal the true parameters and the measured states equal the true states, the (upper bound on) additional cost of our control strategy is zero. In summary, Theorem~\ref{thm: cost} bridges the gap between the error bounds on parameters and states
and the additional cost by implementing the robust control strategy. Lemma~\ref{lem:cost} and Theorem~\ref{thm: cost} answer the additional cost question given by Problem~\ref{Prob-3}.
\vspace{-1.5ex}
\begin{remark}
\label{remark_insight}
    The tightness of the upper bound on the additional cost $\bar{\mathcal{C}}$ depends on the susceptible states $\hat S_{\max}(\hat t_b)$ and $S^*(t^*_h)$. The \baishe{closer} the susceptible proportion at $\hat{t}_h$ \baishe{is} to $1$, the \baishe{smaller} will be the difference between $\hat S_{\max}(\hat t_b)$ and $\hat S_{\max}(\hat t_h)$, as well as $S^*(t^*_h)$ and $S^*(t^*_b)$. Hence, the the upper bound $\bar{\mathcal{C}}$ will become tighter if the change of the susceptible proportion is smaller during the spreading. Additionally, according to~\eqref{eq:add_cost}, smaller estimation error bounds indicated by  $\hat{\beta}_{\max}$ and $\hat{\gamma}_{\min}$  will lead to lower optimality gap $\mathcal{C}$. Thus, it is critical to decrease the parameter estimation error bound, by leveraging the insights from~Theorem~\ref{bound} and Remark~\ref{Remark_Summary}.
\end{remark}

\vspace{-2ex}
\subsection{Impact on Spreading Behavior}
\vspace{-2ex}
After studying the feasibility and additional cost of our proposed robust \update{control} strategy, we explore the spreading behavior under the robust \update{control} strategy to discuss potential benefits and limitations.
\vspace{-1ex}
\begin{coro}
\label{cor:sus}
For any time $t$ up to the herd immunity time step $t^*_h$, $t\in [0,t_h^*]$, the cumulative number of people infected for the optimal \update{control} strategy, $I^*(t)+R^*(t)$, 
will be greater than or equal to the cumulative number of people infected from the proposed \update{control} strategy in Definition~\ref{def:optimal_p}, $I(t)+I(t)$.
\end{coro}
\vspace{-4ex}
\begin{pf}
We first consider the period from $[0,t^*_b]$. 
Based on the fact that $I^*(t)\geq I(t)$ 
for all $t\in [0, t^*_b)$, and based on the proof of Theorem~\ref{lem:inaccu},
\vspace{-2ex}
\begin{align*}
\vspace{-2ex}
    log (S(t))&=log (S(0))-\int_{0}^{t^*_b}(\beta I(\tau)) d\tau \\
    &\geq log (S(0))-\int_{0}^{t^*_b}(\beta I^*(\tau)) d\tau= log (S^*(t)). 
\end{align*} 

\vspace{-1ex}
We further compare $S(t)$ and $S^*(t)$ for all $t\in [t^*_b,t^*_h]$. Following the proof of Lemma~\ref{lem:inaccu}, the optimal control strategy $u^*(t)$ maintains $I^*(t)=\bar{I}$ for all $t\in [t^*_b, t^*_h]$, while the proposed control strategy $\hat{u}(t)$ ensures that $I(t)\leq\bar{I}=I^*(t)$,  for all $t\in [t^*_b, t^*_h]$. Then, from
\vspace{-2ex}
\begin{equation*}
\vspace{-2ex}
    log (S(t))=log (S(t^*_b))-\int_{t^*_b}^{t}(\beta I(\tau)) d\tau,
\end{equation*}
if $ I(t)\leq\bar{I}=I^*(t)$ for all $t\in [t^*_b, t^*_h]$, then $S(t)\geq S^*(t)$, for all $t\in [t^*_b, t^*_h]$ (note that $S(t^*_b)\geq S^*(t^*_b)$). 
Hence, we have shown that $S(t)\geq S^*(t)$ for all $t\in [0, t^*_h]$. Under the fact that $S(t)+I(t)+R(t) = 1$ and $S^*(t)+I^*(t)+R^*(t) = 1$ for all $t\geq 0$, we have that $I(t)+R(t)\leq I^*(t)+R^*(t)$ for all $t\in [0, t^*_h]$. 
Hence, we complete the proof.
\end{pf}
\vspace{-1ex}
Corollary~\ref{cor:sus} shows that the susceptible state dominates the trajectory of the optimal susceptible state for all $t\in[0, t^*_h]$. By overestimating the seriousness of the epidemic, the proposed robust \update{control} strategy can further slow down the infection process, i.e., the cumulative infected population $I(t)+R(t)$. 
Theorem~\ref{lem:inaccu}, Lemma~\ref{lem:cost}, and Corollary~\ref{cor:sus} tackle Problem~\ref{Prob-4}.
From these results, we reach the following conclusions to further answer Problem~\ref{Prob-4}, regarding the benefit and limitation of the robust \update{control} strategy  in Definition~\ref{def:optimal_p}.
\vspace{-1ex}
\begin{remark}
\label{remark}
Compared to the optimal control strategy given in Proposition~\ref{Prop:policy},
the robust \update{control} strategy from Definition~\ref{def:optimal_p} will:
\vspace{-1.5ex}
\begin{enumerate}
    \item Overestimate the seriousness of the epidemic at any given time step;

    \item React earlier to the outbreak and switch back to \update{the zero isolation rate} later;

    \item Cost more or the same in terms of \update{isolation rate} at each time step $t$ for all $t\geq0$;
    \item Be closer to the optimal \update{control} strategy under lower parameter estimation and state measurement error bounds;
    \item 
    Generate \baike{fewer or equal total uninfected individuals \baishe{(including susceptible and recovered individuals)} in the population} at any given time step up to $t^*_h$.
    
\end{enumerate}
\end{remark}
\vspace{-3ex}
\section{Simulations}
\vspace{-2.5ex}
\label{section4}
In this section, we first illustrate the proposed parameter estimation strategy in Section~\ref{sec:2-5}. \baishe{Then, we} demonstrate the proposed robust \update{control} strategy outlined in Definition~\ref{def:optimal_p} through simulations.
\vspace{-2ex}
\subsection{Parameter Estimation Error}
\begin{figure}[h]
    \centering
\includegraphics[ trim = 1cm 0.1cm 2cm 0.1cm, clip, width=\columnwidth]{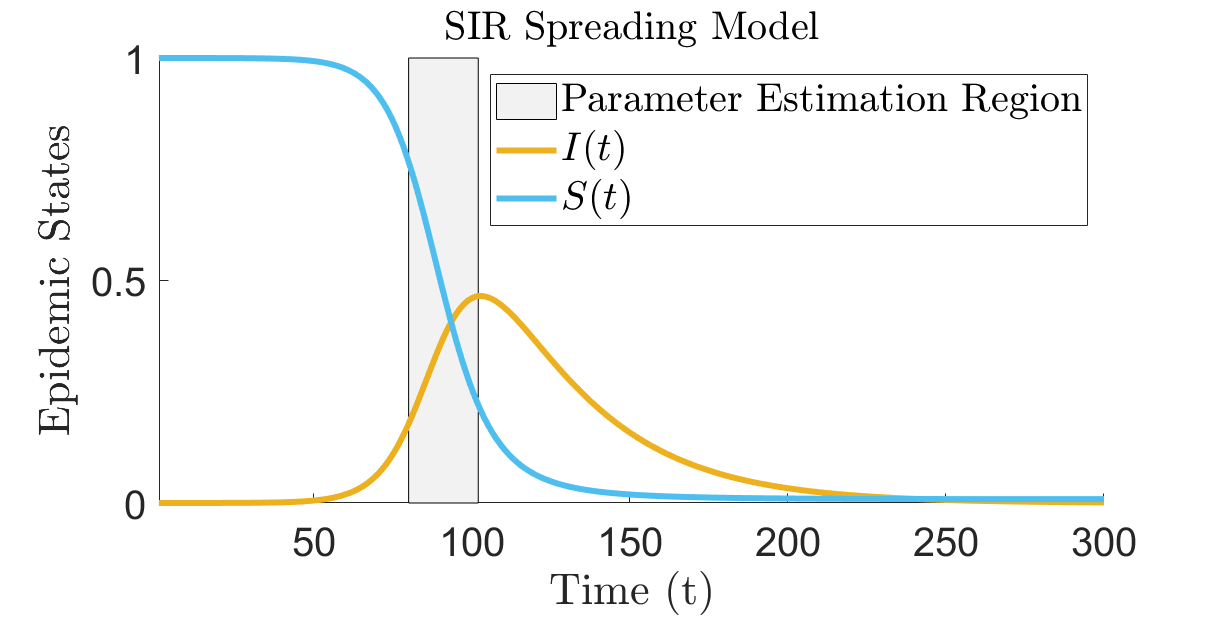}
\centering{}\caption{SIR Spreading Dynamics}
\label{fig:SIR_C}
\end{figure}

\vspace{-2ex}
First, we illustrate the increase in parameter estimation error and the corresponding error bound with respect to the sample step size $h$, as discussed in Lemma~\ref{dt_error} and Theorem~\ref{Error_bound}. Consider an epidemic spreading process captured by an $SIR$ model, with a transmission rate $\beta=0.16$ and removal rate $\gamma=1/30$. To focus on the impact of the sample step size $h$ on the parameter estimation error, we set $u(t)=0$ and $v_{\max}=0$ in~\eqref{touse1}. We set the unit step size as $\bar{h}=0.01$ with the sample step size candidates $h=\alpha\bar{h}$, where $\alpha \in [1,2,3\dots,200]$. We consider the states
from the simulation time step $t=80$ to $t=110$, as illustrated in Figure~\ref{fig:SIR_C}. 
We utilize the states within this range to assess the impact of the sample step size on the parameter estimation error.

\begin{figure}[h]
    \centering
\includegraphics[ trim = 0.4cm 0.1cm  2.5cm 0.1cm, clip, width=\columnwidth]{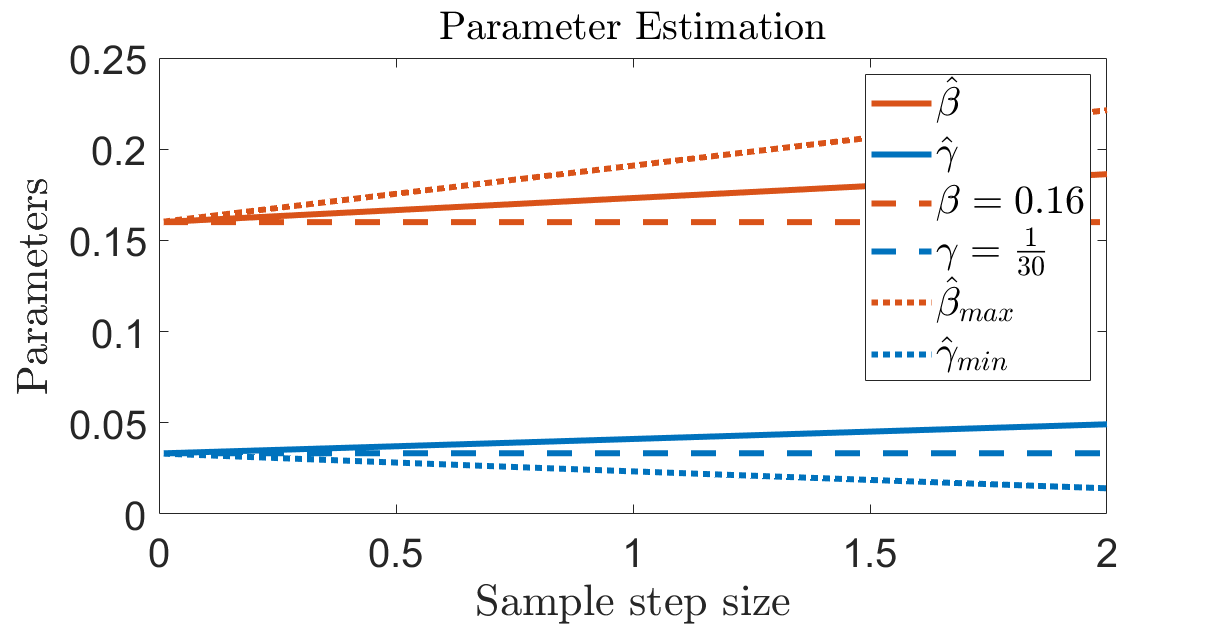}
\centering{}\caption{Parameter Estimation}
\label{fig:para_est}
\end{figure}

\vspace{-1ex}
Following the proposed estimation method in~\eqref{eq:para_est},
we set $x(i)=x(80)$, $x(j)=x(90)$, $x(i+h)=x(80+h)$, and $x(j+h)=x(90+h)$.
Then, we construct $L$ and $Z$ through~\eqref{touse1} and~\eqref{eq:batch}. 
Based on the constructed batch matrices $L$ and $Z$, through~\eqref{eq:para_est}, we generate the estimated model parameters 
$\hat{\Theta} = [\hat\beta \ \ \hat\gamma]$. We plot the estimated model parameters $\hat{\beta}$ and $\hat{\gamma}$ in Figure~\ref{fig:para_est}, which illustrates their variation with respect to the sample step size $h$. 
Furthermore, we observe that the difference between the estimated parameter $\hat{\beta}$ and the true model parameter $\beta$
decreases as the sample step size $h$ decreases. The same observation holds for $\hat{\gamma}$ and $\gamma$.

\begin{figure}[h]
    \centering
\includegraphics[ trim = 0.1cm 0.1cm 2cm 0.1cm, clip, width=\columnwidth]{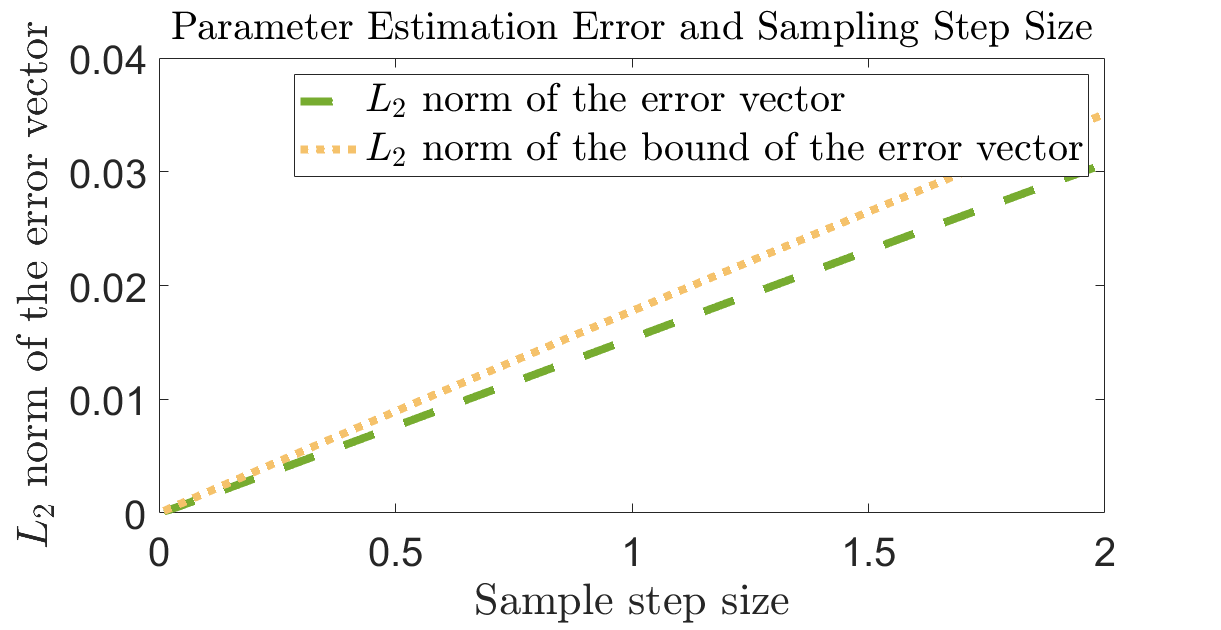}
\centering{}\caption{Estimation Error and Error Bound without Noise}
\label{fig:estimation_norm}
\end{figure}

\vspace{-1ex}
Recall that we assume zero measurement error in this simulation. Therefore, to compute the error bound $b$,  it is sufficient to calculate the first term in ~\eqref{eq:e_bound}.
We compute the parameter
$\lambda_{\min}(ZZ^\top)$  from~\eqref{eq:e_bound} in
Theorem~\ref{Error_bound}. Note that since determining the Lipschitz constant $\zeta$ and $f_{\max}$ requires prior knowledge of the spreading parameters $\beta$ and $\gamma$, we 
set the Lipschitz constant $\zeta=0.055$ through experimentation with the data. 
We plot the upper bound $\hat{\beta}_{\max}$ on the estimated parameter $\hat{\beta}$, and the lower bound $\hat{\gamma}_{\min}$ on the estimated parameter $\hat{\gamma}$ in Figure~\ref{fig:para_est}. Figure~\ref{fig:para_est} further validates Theorem~\ref{Error_bound} by demonstrating the estimation error bounds decreases by using a smaller sample step $h$. 
Figure~\ref{fig:estimation_norm} illustrates that the estimation error bound $\hat{\Theta}$ derived from Theorem~\ref{Error_bound} provides an upper bound on the true estimation error bound $\Theta$. 
Similarly, $\hat{\Theta}$ exhibits a monotonic increase with respect to the sample step size $h$.

\begin{figure}[h]
    \centering
\includegraphics[ trim = 1cm 0.1cm 2.1cm 0.1cm, clip, width=\columnwidth]{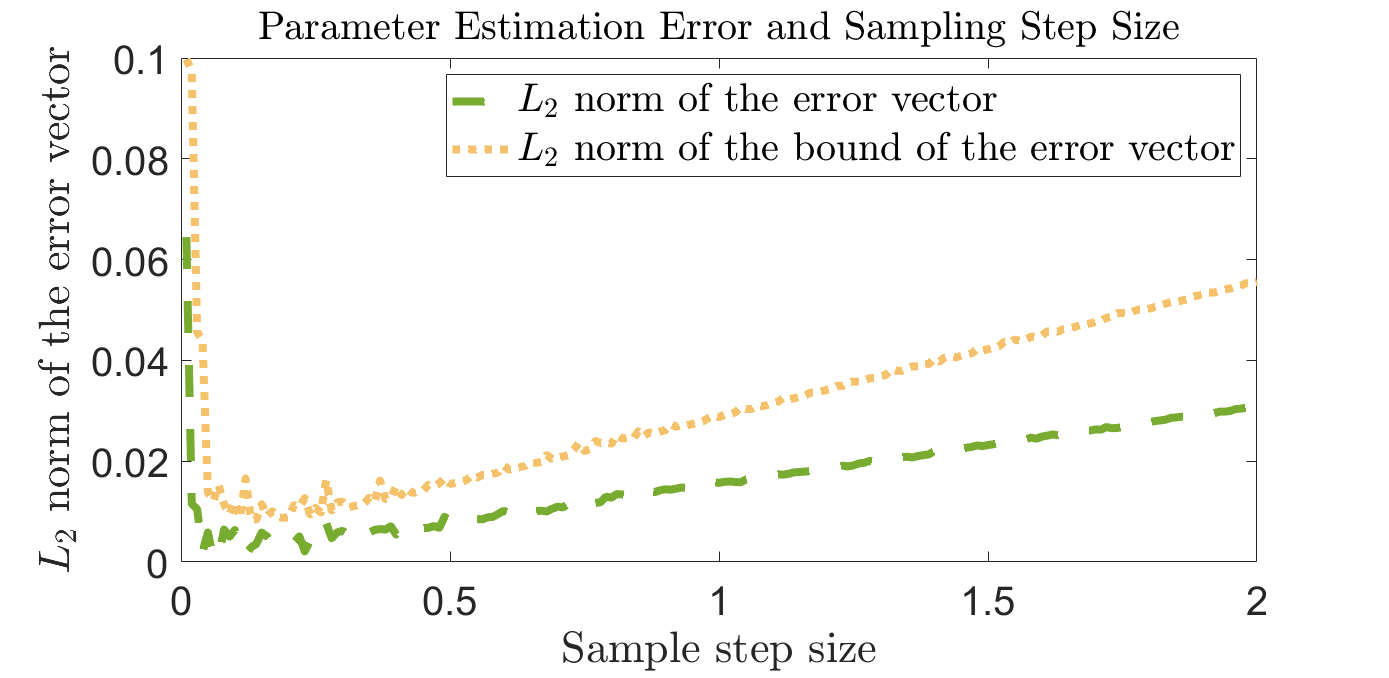}
\centering{}\caption{Estimation Error and Error Bound with Noisy Data}
\label{fig:estimation_norm_noise}
\vspace{-0.5ex}
\end{figure}

\vspace{-0.5ex}
\baishe{We further consider the exact same simulation scenario while adding measurement noise to the spreading data. The signal-to-noise ratio is given by $100dB$. We show through Figure~\ref{fig:estimation_norm_noise} that, compared to Figure~\ref{fig:estimation_norm}, the upper bound on the estimation error bound $\hat\Theta$, derived from Theorem~\ref{Error_bound}, and the true error bound $\Theta$ decrease with respect to the sample step size $h$ first, then exhibit a monotonic increase with respect to the sample step size $h$. This result illustrates the second and third terms in Equation~\eqref{eq:e_bound}, such that when we have noisy measurement and when the sampling step size is too small, we will have higher estimation error.}

\vspace{-2ex}
\subsection{Overestimation Vs. Underestimation}
\vspace{-2.5ex}
Consider an epidemic spreading process described by~\eqref{Eq: Con_Dynamics} with $\beta=0.16$ and $\gamma =1/30$. 
The objective is to minimize the \update{cumulative isolation rate} during the epidemic, as given by~\eqref{eq:prob}, while maintaining the infection level at or below $1\%$ of the population, i.e., $\bar{I}=0.01$. The upper bound on the daily isolation rate is $\bar{u}=15\%$. 
The initial conditions are $I(0)=0.00001$, $R(0)=0$, and $S(0)=1-I(0)$. The measured states are corrupted by noise, and the signal-to-noise ratio is $55dB$. 
We leverage the \update{control} policies provided by Proposition~\ref{Prop:policy} and Definition~\ref{def:optimal_p},  and then compare the results.

\begin{figure*}
 \centering
\includegraphics[trim = 2cm 0.5cm 4cm 1cm, clip,width=1.0\textwidth]{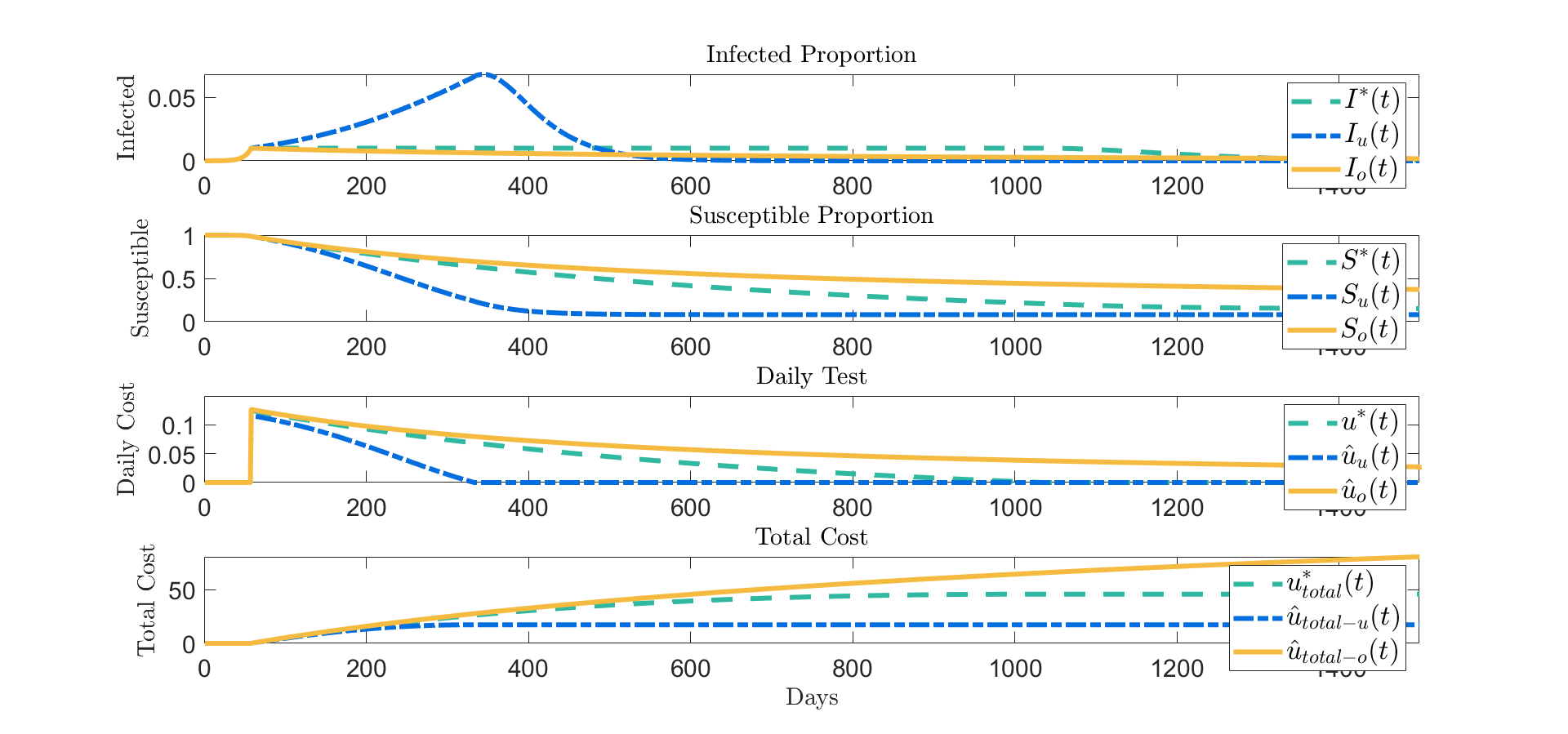}
\caption{Comparison Between \update{Control} Strategies. \baishe{Figure~\ref{fig:testing_comp} captures three scenarios: the green lines represent the spreading process and the cost under the optimal control strategy; the yellow lines depict the spreading process and the cost when overestimating the severity of the spread; and the blue lines illustrate the spreading process and the cost when underestimating the spread. As demonstrated by the example, underestimating the spreading process can cause significant outbreaks, as indicated by the peak infection, $I_u(t)$.  Overestimating the severity of the epidemic may lead to higher \update{isolation} costs, captured by $\hat{u}_o(t)$ and $\hat{u}_{total-o}(t)$. However, as studied in this work, overestimating the severity of a spreading process while employing a 
\update{control}
strategy enhances the robustness of the strategy against model uncertainty arising from inaccurate parameters.}}
\label{fig:testing_comp}
\end{figure*}

\vspace{-1.5ex}
In addition to the optimal \update{control} strategy ($u^*(t)$) that leverages the true parameters and states, we consider two other \update{control}  strategies. The first strategy $u_u(t)$  implements  the measured states and the underestimated parameters \update{in the optimal control strategy given by Proposition~\ref{Prop:policy}}. The second strategy $\hat u_o(t)$  follows Definition~\ref{def:optimal_p}, where the spreading parameters and states are overestimated. Figure~\ref{fig:testing_comp} provides a comparison of the epidemic dynamics under these three \update{control} strategies: $u^*(t)$, $\hat u_u(t)$, and $\hat u_o(t)$. 
We begin by examining the consequences of underestimating and overestimating the spreading process when implementing these \update{control} strategies.
Note that we use $I^*(t)$, $S^*(t)$ to denote the system trajectories under the optimal \update{isolation}  rate $u^*(t)$ and the cumulative cost $u_{total}^*(t)$. Similarly, $I_u(t)$, $S_u(t)$ and $I_o(t)$, $S_o(t)$ represent the true system trajectories under the underestimated strategy $\hat{u}_u(t)$ and the overestimated strategy $\hat u_o(t)$, respectively. \bshee{The corresponding measured states ($\hat I_u(t)$, $\hat S_u(t)$, and $\hat I_o(t)$, $\hat S_o(t)$) which we leverage for parameter estimation and control design are not shown.}

\vspace{-1.5ex}
The trajectories in Figure~\ref{fig:testing_comp} demonstrate that the control system is feasible when overestimating the spreading parameters and states, as illustrated by $I_o(t)$ in Figure~\ref{fig:testing_comp}. However, when underestimating the spreading process by directly employing the optimal control strategy from Proposition~\ref{Prop:policy} ($\hat{u}_u(t)$), the system becomes infeasible, as shown in Figure~\ref{fig:testing_comp}, by infection state $I_u(t)$ continuing to increase even after reaching the threshold $\bar{I}$. This phenomenon arises because when the \update{control} strategy transitions from Stage 1 to Stage 2 in Proposition~\ref{Prop:policy}, the \update{isolation} rate with the underestimated parameters and states will result in the underestimation of the seriousness of the epidemic. Consequently, the \update{control} strategy will generate \baishe{an} insufficient \update{isolation} rate $\hat{u}_u(t)$ that fails to maintain the infection level below the threshold $\bar{I}$. 
Recall from Lemma~\ref{lem:cost}, the optimal control policy gives the pointwise smallest \update{isolation rate} to ensure the system is feasible. Hence, the condition $\hat{u}_u(t)\leq u^*(t)$ during the outbreak will result in the system becoming infeasible. 

\vspace{-1.5ex}
Regarding the second statement of Remark~\ref{remark}, the simulation reveals that the system under $\hat u_o(t)$ takes longer to reach herd immunity compared to the system under the optimal \update{control} strategy $u^*(t)$. The \update{control} strategy $\hat u_o(t)$  generated from overestimating the spreading process,  results in higher \update{isolation} rates than the optimal strategy pointwise, i.e., $\hat u_o(t)\geq u^*(t)$, for all $t\geq 0$.
By comparing the susceptible states, it can be observed that $\hat u_o(t)$ leads to an equal or smaller total uninfected population at any given time step, i.e., $S_o(t)\geq S^*(t)$, for all $t\geq0$. This result
reflects Corollary~\ref{cor:sus} \baishe{in} that the \update{control} strategy $\hat u_o(t)$ will cause fewer people to be infected over the course of the outbreak, that is, $I^*(t)+R^*(t)\geq I_o(t)+R_o(t)$ for all $t\in [0, t^*_h]$.

\section{Conclusion}
\vspace{-2ex}
\label{section5}
In this work, we \baishe{focused} on studying the impact of uncertainties caused by parameter estimation and measurement error on optimal epidemic mitigation. 
We \baishe{presented} a method to analyze the parameter estimation error bounds resulting from the sampling process of the continuous-time $SIR$ spreading model. We \baishe{found} that the parameter estimation error is determined by the spreading dynamics, the sampling time interval, and the measurement error. Further,
we \baishe{demonstrated} the effectiveness of our proposed robust \update{control} strategy when overestimating the severity of the epidemic. We \baishe{investigated} how the parameter estimation and measurement error bounds influence the robust \update{control} strategy. Compared to the optimal \update{control} strategy, the robust \update{control} strategy can effectively flatten the curve, albeit at a higher cost that is determined by the tightness of the error bounds. In addition, we analytically \baishe{proved} that our strategy yields an equal or smaller cumulative number of infected individuals at any given time step until reaching optimal herd immunity.
We assume that the parameter estimation process is performed offline and that the dynamics of the epidemic are time-invariant and deterministic. Future research will focus on developing parameter estimation strategies for time-varying, stochastic spreading models to better capture real-world epidemic dynamics.

\bibliographystyle{unsrt}        
\vspace{-1ex}
\bibliography{autosam}

\begin{thebibliography}{10}

\bibitem{tsay2020modeling}
Calvin Tsay, Fernando Lejarza, Mark~A Stadtherr, and Michael Baldea.
\newblock Modeling, state estimation, and optimal control for the {US COVID-19} outbreak.
\newblock {\em Scientific Reports}, 10(1):1--12, 2020.

\bibitem{perkins2020optimal}
T~Alex Perkins and Guido Espa{\~n}a.
\newblock Optimal control of the {COVID-19} pandemic with non-pharmaceutical interventions.
\newblock {\em Bulletin of Mathematical Biology}, 82(9):1--24, 2020.

\bibitem{acemoglu2021optimal}
Daron Acemoglu, Alireza Fallah, Andrea Giometto, Daniel Huttenlocher, Asuman Ozdaglar, Francesca Parise, and Sarath Pattathil.
\newblock Optimal adaptive testing for epidemic control: {Combining} molecular and serology tests.
\newblock {\em Automatica}, 160:111391, 2024.

\bibitem{morris2021optimal}
Dylan~H Morris, Fernando~W Rossine, Joshua~B Plotkin, and Simon~A Levin.
\newblock Optimal, near-optimal, and robust epidemic control.
\newblock {\em Communications Physics}, 4(1):1--8, 2021.

\bibitem{martins2023n}
Douglas Martins, Amit Bhaya, and Fernando Pazos.
\newblock N-step-ahead optimal control of a compartmental model of {COVID-19}.
\newblock {\em Journal of Control, Automation and Electrical Systems}, pages 1--15, 2023.

\bibitem{van2023effective}
Klaske van Heusden, Greg~E Stewart, Sarah~P Otto, and Guy~A Dumont.
\newblock Effective pandemic policy design through feedback does not need accurate predictions.
\newblock {\em PLOS Global Public Health}, 3(2):e0000955, 2023.

\bibitem{kohler2020robust}
Johannes K{\"o}hler, Lukas Schwenkel, Anne Koch, Julian Berberich, Patricia Pauli, and Frank Allg{\"o}wer.
\newblock Robust and optimal predictive control of the {COVID-19} outbreak.
\newblock {\em Annu. Rev. in Contr.}, 2020.

\bibitem{carli2020model}
Raffaele Carli, Graziana Cavone, Nicola Epicoco, Paolo Scarabaggio, and Mariagrazia Dotoli.
\newblock Model predictive control to mitigate the {COVID-19} outbreak in a multi-region scenario.
\newblock {\em Annu. Rev. in Contr.}, 50:373--393, 2020.

\bibitem{zino2021analysis}
Lorenzo Zino and Ming Cao.
\newblock Analysis, prediction, and control of epidemics: A survey from scalar to dynamic network models.
\newblock {\em IEEE Circuits and Systems Magazine}, 21(4):4--23, 2021.

\bibitem{she2022mpcepi}
Baike She, Shreyas Sundaram, and Philip~E Par\'e.
\newblock A learning-based model predictive control framework for real-time {SIR} epidemic mitigation.
\newblock In {\em Proc. of the American Contr. Conf.}, pages 2565 --2570, 2022.

\bibitem{khadilkar2020optimising}
Harshad Khadilkar, Tanuja Ganu, and Deva~P Seetharam.
\newblock Optimising lockdown policies for epidemic control using reinforcement learning.
\newblock {\em Trans. of the Indian National Aca. of Engine.}, 5(2):129--132, 2020.

\bibitem{scarabaggio2021nonpharmaceutical}
Paolo Scarabaggio, Raffaele Carli, Graziana Cavone, Nicola Epicoco, and Mariagrazia Dotoli.
\newblock Nonpharmaceutical stochastic optimal control strategies to mitigate the {COVID-19} spread.
\newblock {\em IEEE Trans. on Autom. Sci. and Engine.}, 2021.

\bibitem{mubarak2022individual}
Mohammad Mubarak, James Berneburg, and Cameron Nowzari.
\newblock Individual non-pharmaceutical intervention strategies for stochastic networked epidemics.
\newblock In {\em Proc. IEEE 61st Conference on Decision and Control (CDC)}, pages 5627--5632. IEEE, 2022.

\bibitem{bastani2021efficient}
Hamsa Bastani, Kimon Drakopoulos, Vishal Gupta, Ioannis Vlachogiannis, Christos Hadjicristodoulou, Pagona Lagiou, Gkikas Magiorkinis, Dimitrios Paraskevis, and Sotirios Tsiodras.
\newblock Efficient and targeted {COVID-19} border testing via reinforcement learning.
\newblock {\em Nature}, 599(7883):108--113, 2021.

\bibitem{bloem2009optimal}
Michael Bloem, Tansu Alpcan, and Tamer Ba{\c{s}}ar.
\newblock Optimal and robust epidemic response for multiple networks.
\newblock {\em Control Engineering Practice}, 17(5):525--533, 2009.

\bibitem{nowzari2016epidemics}
Cameron Nowzari, Victor~M Preciado, and George~J Pappas.
\newblock {Analysis and control of epidemics: A survey of spreading processes on complex networks}.
\newblock {\em IEEE Control Systems Magazine}, 36(1):26--46, 2016.

\bibitem{di2017optimal}
Paolo Di~Giamberardino and Daniela Iacoviello.
\newblock Optimal control of {SIR} epidemic model with state dependent switching cost index.
\newblock {\em Biomedical Signal Processing and Control}, 31:377--380, 2017.

\bibitem{sharomi2017optimal}
Oluwaseun Sharomi and Tufail Malik.
\newblock Optimal control in epidemiology.
\newblock {\em Annals of Operations Research}, 251(1-2):55--71, 2017.

\bibitem{di2019optimal}
Paolo Di~Giamberardino and Daniela Iacoviello.
\newblock Optimal resource allocation to reduce an epidemic spread and its complication.
\newblock {\em Information}, 10(6):213, 2019.

\bibitem{liu2019bivirus}
Ji~Liu, Philip~E Par{\'e}, Angelia Nedi\'c, Choon~Yik Tang, Carolyn~L Beck, and Tamer Ba\c{s}ar.
\newblock Analysis and control of a continuous-time bi-virus model.
\newblock {\em IEEE Trans. on Autom. Contr.}, 64(12):4891--4906, 2019.

\bibitem{dangerfield2019resource}
Ciara~E Dangerfield, Martin Vyska, and Christopher~A Gilligan.
\newblock Resource allocation for epidemic control across multiple sub-populations.
\newblock {\em Bulletin of Mathematical Biology}, 81(6):1731--1759, 2019.

\bibitem{preciado2014epidemic_optimal}
Victor~M Preciado, Michael Zargham, Chinwendu Enyioha, Ali Jadbabaie, and George Pappas.
\newblock Optimal resource allocation for network protection: A geometric programming approach.
\newblock {\em IEEE Transactions on Control of Network Systems}, 1(1):99--108, 2014.

\bibitem{han2015data}
Shuo Han, Victor~M Preciado, Cameron Nowzari, and George~J Pappas.
\newblock Data-driven network resource allocation for controlling spreading processes.
\newblock {\em IEEE Trans. on Netw. Sci. and Engine.}, 2(4):127--138, 2015.

\bibitem{chowell2017fitting}
Gerardo Chowell.
\newblock Fitting dynamic models to epidemic outbreaks with quantified uncertainty: {A} primer for parameter uncertainty, identifiability, and forecasts.
\newblock {\em Infec. Dise. Model.}, 2(3):379--398, 2017.

\bibitem{baker2018mechanistic}
Ruth~E Baker, Jose-Maria Pena, Jayaratnam Jayamohan, and Antoine J{\'e}rusalem.
\newblock Mechanistic models versus machine learning, a fight worth fighting for the biological community?
\newblock {\em Bio. Lett.}, 14(5):20170660, 2018.

\bibitem{wilke2020predicting}
Claus~O Wilke and Carl~T Bergstrom.
\newblock Predicting an epidemic trajectory is difficult.
\newblock {\em PNAS}, 117(46):28549--28551, 2020.

\bibitem{casella2020can}
Francesco Casella.
\newblock Can the {COVID-19} epidemic be controlled on the basis of daily test reports?
\newblock {\em IEEE Contr. Sys. Let.}, 5(3):1079--1084, 2020.

\bibitem{sontag2023explicit}
Eduardo~D Sontag.
\newblock An explicit formula for minimizing the infected peak in an {SIR} epidemic model when using a fixed number of complete lockdowns.
\newblock {\em International Journal of Robust and Nonlinear Control}, 33(9):4708--4731, 2023.

\bibitem{olejarz2023optimal}
Jason~W Olejarz, Kirstin I~Oliveira Roster, Stephen~M Kissler, Marc Lipsitch, and Yonatan~H Grad.
\newblock Optimal environmental testing frequency for outbreak surveillance.
\newblock {\em Epidemics}, page 100750, 2024.

\bibitem{kirshner2014unique}
Hagai Kirshner, Michael Unser, and John~Paul Ward.
\newblock On the unique identification of continuous-time autoregressive models from sampled data.
\newblock {\em IEEE Transactions on Signal Processing}, 62(6):1361--1376, 2014.

\bibitem{zhu2022performance}
Xinghua Zhu, Die Gan, and Zhixin Liu.
\newblock Performance analysis of least squares of continuous-time model based on sampling data.
\newblock {\em IEEE Control Systems Letters}, 6:3086--3091, 2022.

\bibitem{grundel2021coordinate}
Sara~M Grundel, Stefan Heyder, Thomas Hotz, Tobias K~S Ritschel, Philipp Sauerteig, and Karl Worthmann.
\newblock How to coordinate vaccination and social distancing to mitigate {SARS-CoV-2} outbreaks.
\newblock {\em SIAM Journal on Applied Dynamical Systems}, 20(2):1135--1157, 2021.

\bibitem{she2023reverse}
Baike She, Rebecca~Lee Smith, Ian Pytlarz, Shreyas Sundaram, and Philip~E Par{\'e}.
\newblock Reverse engineering the reproduction number: {A} framework for data-driven counterfactual analysis, strategy evaluation, and feedback control of epidemics.
\newblock {\em arXiv preprint arXiv:2311.01471}, 2023.

\bibitem{she2022optimal}
Baike She, Shreyas Sundaram, and Philip~E Par{\'e}.
\newblock Optimal mitigation of {SIR} epidemics under model uncertainty.
\newblock In {\em Proc. IEEE 61st Conf. on Deci. and Cont. (CDC)}, pages 4333--4338. IEEE, 2022.

\bibitem{kermack1927_sir}
William~O. Kermack and Anderson~G. McKendrick.
\newblock A contribution to the mathematical theory of epidemics.
\newblock {\em Proceedings of the Royal Society A}, 115(772):700--721, 1927.

\bibitem{draper1998applied}
Norman~R Draper and Harry Smith.
\newblock {\em Applied Regression Analysis}, volume 326.
\newblock John Wiley \& Sons, 1998.

\bibitem{roosa2019assessing}
Kimberlyn Roosa and Gerardo Chowell.
\newblock Assessing parameter identifiability in compartmental dynamic models using a computational approach: {Application} to infectious disease transmission models.
\newblock {\em Theoretical Biology and Medical Modelling}, 16:1--15, 2019.

\bibitem{zimmer2017likelihood}
Christoph Zimmer, Reza Yaesoubi, and Ted Cohen.
\newblock A likelihood approach for real-time calibration of stochastic compartmental epidemic models.
\newblock {\em PLoS Computational Biology}, 13(1):e1005257, 2017.

\bibitem{morato2020optimal}
Marcelo~M Morato, Saulo~B Bastos, Daniel~O Cajueiro, and Julio~E Normey-Rico.
\newblock An optimal predictive control strategy for {COVID-19 (SARS-CoV-2)} social distancing policies in {Brazil}.
\newblock {\em Annu. Rev. in Contr.}, 50:417--431, 2020.

\bibitem{peni2020nonlinear}
Tam{\'a}s P{\'e}ni, Bal{\'a}zs Csutak, G{\'a}bor Szederk{\'e}nyi, and Gergely R{\"o}st.
\newblock Nonlinear model predictive control with logic constraints for {COVID-19} management.
\newblock {\em Nonlinear Dynamics}, 102(4):1965--1986, 2020.

\bibitem{grundel2020much}
Sara Grundel, Stefan Heyder, Thomas Hotz, Tobias~KS Ritschel, Philipp Sauerteig, and Karl Worthmann.
\newblock How much testing and social distancing is required to control {COVID-19}? {Some} insight based on an age-differentiated compartmental model.
\newblock {\em SIAM Journal on Control and Optimization}, 60(2):S145--S169, 2022.

\bibitem{sereno2021model}
Juan~Esteban Sereno, Agustina D’Jorge, Antonio Ferramosca, Esteban~A Hernandez-Vargas, and Alejandro~H Gonzalez.
\newblock Model predictive control for optimal social distancing in a type {SIR-switched model}.
\newblock {\em Ifac-papersonline}, 54(15):251--256, 2021.

\bibitem{mei2017epidemics_review}
Wenjun Mei, Shadi Mohagheghi, Sandro Zampieri, and Francesco Bullo.
\newblock On the dynamics of deterministic epidemic propagation over networks.
\newblock {\em Annu. Rev. in Contr.}, 44:116--128, 2017.

\bibitem{malik2013discretization}
Fahad~Mumtaz Malik, Mohammad~Bilal Malik, and Khalid Munawar.
\newblock Discretization error bounds for sampled-data control of nonlinear systems.
\newblock {\em Arabian Journal for Science and Engineering}, 38:3429--3435, 2013.

\bibitem{pare2018analysis}
Philip~E Par{\'e}, Ji~Liu, Carolyn~L Beck, Barrett~E Kirwan, and Tamer Ba{\c{s}}ar.
\newblock Analysis, estimation, and validation of discrete-time epidemic processes.
\newblock {\em IEEE Transactions on Control Systems Technology}, 28(1):79--93, 2018.

\bibitem{fine1993herd}
Paul~EM Fine.
\newblock Herd immunity: {History,} theory, practice.
\newblock {\em Epidemiologic Reviews}, 15(2):265--302, 1993.

\end{thebibliography}

 




\end{document}